\pdfoutput=1
\documentclass[8pt]{article}  \usepackage{times}
\usepackage{graphicx}

\usepackage{color}

\begin{document}

\twocolumn[{\LARGE \textbf{Review: Lipid Ion Channels\\*[0.0cm]}}

{\large Thomas Heimburg}\\
{\small Membrane Biophysics Group, Niels Bohr Institute, University of Copenhagen, Denmark}\\

{{\normalsize \textbf{ABSTRACT\hspace{0.5cm} The interpretation electrical phenomena in biomembranes is usually based on the assumption that the experimentally found discrete ion conduction events are due to a particular class of proteins called ion channels while the lipid membrane is considered being an inert electrical insulator. The particular protein structure is thought to be related to ion specificity, specific recognition of drugs by receptors and to macroscopic phenomena as nerve pulse propagation. However, lipid membranes in their chain melting regime are known to be highly permeable to ions, water and small molecules, and are therefore not always inert. In voltage-clamp experiments one finds quantized conduction events through protein-free membranes in their melting regime similar to or even undistinguishable from those attributed to proteins. This constitutes a conceptual problem for the interpretation of electrophysiological data obtained from biological membrane preparations.\\
Here, we review the experimental evidence for lipid ion channels, their properties and the physical chemistry underlying their creation. We introduce into the thermodynamic theory of membrane fluctuations from which the lipid channels originate. Furthermore, we demonstrate how the appearance of lipid channels can be influenced by the alteration of the thermodynamic variables (temperature, pressure, tension, chemical potentials) in a coherent description that is free of parameters. This description leads to pores that display dwell times closely coupled to the fluctuation lifetime via the fluctuation-dissipation theorem. Drugs as anesthetics and neurotransmitters are shown to influence the channel likelihood and their lifetimes in a predictable manner. We also discuss the role of proteins in influencing the likelihood of lipid channel formation.
}\\*[0.0cm] }}
]

\footnotesize { \textbf{keywords:} lipid pores, phase transitions, fluctuations, anesthetics, receptors, channel proteins, fluctuation-dissipation, electroporation}

\renewcommand{\topfraction}{0.95} 
\renewcommand{\textfraction}{0.05} 
\renewcommand{\floatpagefraction}{0.95} 

\normalsize
\section{Introduction}\label{Intro}
Biological membranes mainly consist of a lipid bilayer into which proteins are embedded. The mass (or volume) ratio between proteins and lipids ranges between 0.25 (lung surfactant) and 4 (purple membrane of halobacteria). Typical, biomembranes display a protein-lipid ratio of approximately one. This includes the extra-membraneous parts of the proteins and surface-associated protein. Therefore, the intra-membrane parts of the proteins represent a smaller fraction of the central part of the membrane. Thus, even in densely crowded biological membranes most of the in-plane membrane area consists of lipids. \\
Textbook models about the electrical transport properties of membranes consider the lipid bilayer as an electrical insulator. For instance, the Hodgkin- Huxley model of the nervous impulse considers the membrane as an inert capacitor and attributes currents to conductance changes occurring in ion selective protein channels \cite{Hodgkin1952}. The activity of protein ion channels is associated to quantized current events of order of 1-50 pA (pico Amp\`eres) at a voltage of order 100mV corresponding to channel conductances of about 10-500 pS (pico Siemens). The typical open-dwell time of such channels is of the order of a few milliseconds.\\ 
Interestingly, however, synthetic lipid membranes close to phase transitions are not inert but are very permeable for small molecules \cite{Blicher2009}, ions \cite{Papahadjopoulos1973, Nagle1978b, Sabra1996, Blicher2009} and water \cite{Jansen1995}. Further, they display stepwise conductance events very similar to those of proteins, i.e., with similar conductances and life times (some early publications:  \cite{Yafuso1974, Antonov1980, Boheim1980, Kaufmann1983a, Kaufmann1983b, Antonov1985}). In their appearance, they are practically indistinguishable from recordings of protein-containing membranes. Although this is surprisingly little known, it constitutes a severe problem for the investigation of biological membranes and protein channels. These events will be called lipid channels throughout this manuscript. The finding of an enhanced lipid membrane permeability close to transitions is particularly important since biomembranes seem to exist in a state slightly (about 10-15$^{\circ}$) above melting transitions. The membrane compositions of fish and bacteria adapt upon changes in body temperature, pressure or solvent conditions such that this membrane state is maintained (reviewed in \cite{Heimburg2007a}). It seems therefore plausible to assuming a functional purpose for preserving a particular physical state of the membrane. \\ 
The most common ways to measure conductances of membranes are the black lipid membrane (BLM) technique by Montal \& Mueller \cite{Montal1972} and the patch-clamp technique pioneered by Neher \& Sackmann \cite{Neher1976}.  In the Montal-Mueller technique an artificial membrane is formed over a small teflon hole (normally with a diameter of 50-250 $\mu$m). The patch-clamp technique relies on the measurement of currents through small membrane patches defined by a glass capillary of a diameters ranging from 1-10 $\mu$m. While the Montal-Mueller technique is suited for measuring artificial systems of defined composition, e.g., one single protein species reconstituted into a single lipid bilayer, the patch-clamp method is mainly used for recording currents through biological membranes with complex and often unknown composition. Patch pipettes have also been used for measurements on synthetic BLMs \cite{Coronado1983, Hanke1984}. In both techniques, the observed membrane area is much larger than that of a protein. The diameter of an ion channel protein, e.g. the potassium channel, is about 5nm. This implies that the smallest patch in a patch-clamp experiment is already 40000 times larger than the cross-sectional area of the protein. The hole in an BLM experiment is rather 400 million time larger than a typical protein channel area (assuming an aperture of 100 $\mu$m). However, the electrophysiological experiment in itself cannot tell where along which path the ion currents flow. Thus, if one wants to attribute a particular current event to a protein one has to assume that at least 99.998\% of the surrounding membrane (including the other membrane proteins) are inert in the electrophysiological experiment. Therefore, it can generally not be excluded on the basis of the experiment itself that during an electrophysiological experiments one also finds currents through the lipid membrane. In the Hodgkin-Huxley paper \cite{Hodgkin1952}, the possibility of leak currents is included. If these are structureless small amplitude currents in the background this does not represent a major problem. However, leak currents represent a significant conceptual problem to protein channel recordings if they display a similar signature as the protein events.  \\
The purpose of this review is to demonstrate that in fact the ion currents through lipid membranes resemble those through proteins to a degree that they become indistinguishable. Their occurrence depends on temperature, lateral tension, membrane-associated drugs as anesthetics and neurotransmitters, pH, calcium concentration, voltage and other thermodynamic variables. Here, we review the literature on membrane permeability and lipid ion channels, and introduce into the thermodynamics of their creation. In this context we discuss a possible role for proteins as membrane perturbations that alter the state of the lipid membrane.

\section{Macroscopic changes of permeability in transitions}\label{MacroscopicExp}
\subsection{Experiments}
During the 1960s and 70s, many researchers were investigating the permeability of biomembranes for ions. Proteins were assumed being the major players. However, accurate controls of the permeability of the lipid membrane for ions was needed as a control. Hauser et al. \cite{Hauser1972} studied the permeability of lipid extracts from egg yolk or ox brain and found the permeability for ions being small compared to the  permeability of biological membranes. The authors concluded that the role of the lipid bilayer for permeability is negligible for the case of a biological membrane suggesting that the predominant part of permeability has to be attributed to proteins.\\
\begin{figure}[htb!]
    \begin{center}
	\includegraphics[width=8.5cm]{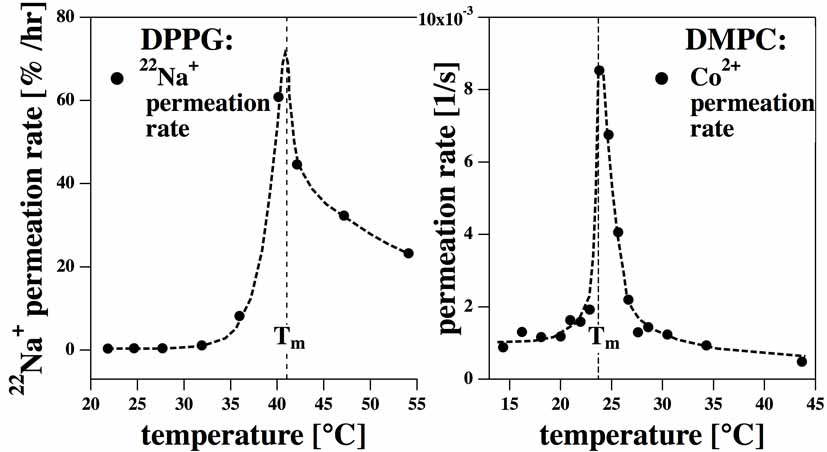}
	\parbox[c]{8cm}
{ \caption{\textit{Left: Permeability of DPPG vesicles for radioactive sodium, $^{22}Na^+$, close to the melting transition, adapted from \cite{Papahadjopoulos1973}. Right: Permeability of DMPC membranes for Co$^2+$. Courtesy O.G. Mouritsen, cf. \cite{Sabra1996}. The dotted profiles are a gide to the eye. The dotted vertical line marks the melting transition temperature, T$_m$.}
	\label{Figure1a}}}
    \end{center}
\end{figure}
However, it was soon shown that the permeability can be orders of magnitude larger close to the chain melting regime of lipid membranes. Papahadjopoulos et al. \cite{Papahadjopoulos1973} were the first to demonstrate that the permeability for sodium ions (they used radiolabeled $^{22}$Na$^{+}$ ions) increased by at least a factor of 100 in the phase transitions of dipalmitoyl phosphatidylglycerol (DPPG) and dipalmitoyl phosphatidylcholine (DMPC) in agreement with the phase transitions of these lipids as measured by fluorescence changes of embedded markers. The permeability curve for DPPG membranes is shown in Fig. \ref{Figure1a} (left). The permeation time scales in this paper seem extremely small (of the order of hours) which might be related to the fact that none of the data points was recorded at the phase transition directly. The permeation profile for DPPC was found to be similar similar (not shown). It was demonstrated that cholesterol both abolishes the permeability maximum and the chain melting discontinuity. Along the same lines Sabra et al. \cite{Sabra1996} found that the permeability of dimyristoyl phophatidylcholine (DMPC) membranes for Co$^{2+}$ was drastically enhanced in the phase transition regime (Fig. \ref{Figure1a}, right). These authors also demonstrated that the insecticide lindane changes the permeability. This phenomenon will be discussed in more detail in section \ref{Variables.3}.\\
Jansen and collaborators \cite{Jansen1995} showed that membranes in their transition are much more permeable to water (Fig. \ref{Figure1b}, left). Vesicles filled with D$_2$O display a contrast with respect to an H$_2$O background leading to enhanced light scattering in an optical experiment. Permeation for D$_2$O was monitored in a rapid mixing stopped-flow experiment. It leads to a mixing of H$_2$O and D$_2$O and a loss of scattering contrast. The authors found that the permeability of a DMPC membrane changes strongly in the phase transition regime such that exchange of water from a vesicle is complete after 2ms, the fastest time that could be recorded in this experiment.  The same finding was reported for other lipids as DPPC and distearoyl phosphatidylcholine (DSPC), a series of phosphatidyl ethanolamines, phosphatidyl serines and phosphatidylglycerols, rendering this study one of the most complete works on membrane permeability. Since water permeability typically is discussed in the context of aquaporines \cite{Agre1998} it is important to point out the the membrane itself is highly permeable to water once one is close to a transition, which seems to be the case for many biological membranes \cite{Heimburg2005c, Heimburg2007a}.  
\begin{figure}[htb!]
    \begin{center}
	\includegraphics[width=9cm]{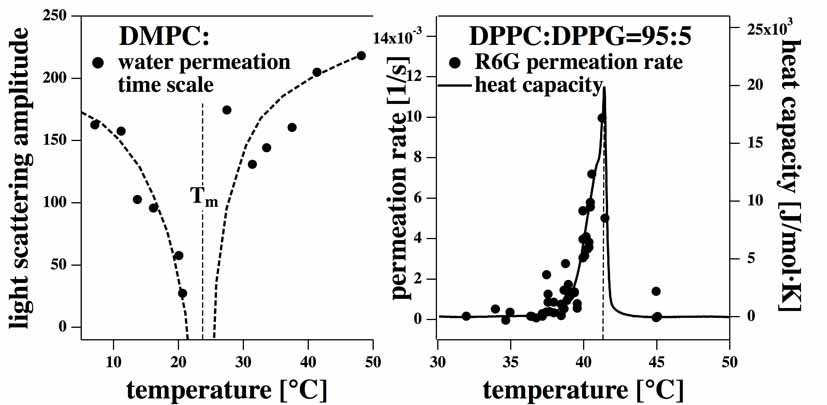}
	\parbox[c]{8cm}
{ \caption{\textit{Left: Permeation time scale for water. Light scattering of DMPC vesicles filled with D$_2$O in a H$_2$O environment is monitored after a fixed time in a stopped-flow experiment. The decrease of the scattering amplitude close to T$_m$=23.7$^{\circ}$C is due to fast mixing of D$_2$O and H$_2$O. Adapted from \cite{Jansen1995}. The dotted line is a guide to the eye. Right: The permeation rate of Rhodamine 6G out of vesicles from a mixture of DPPC and DPPG (95:5) in the presence of 200mM NaCl. The solid line is the experimental heat capacity profile for vesicles of the identical preparation. Adapted from \cite{Blicher2009}.}
	\label{Figure1b}}}
    \end{center}
\end{figure}
Blicher and coworkers \cite{Blicher2009} found in fluorescence correlation spectroscopy measurements that vesicles mixtures of DPPC and DPPG became significantly more permeable for the positively charged fluorescence marker rhodamine 6G (R6G) in the phase transition regime (Fig. \ref{Figure1b}, right). While the permeation time scale in the solid lipid phase was of the order of many hours, it was 100 seconds and faster in the transition regime (the fastest time scale that could be recorded in the FCS experiment). The characteristic dimension of this marker is approximately 1nm. A control experiment with a fluorescent-labeled sugar that are larger than R6G did not show any measurable permeation on the timescale of hours. This demonstrates that 
\begin{enumerate}
  \item vesicles stay intact and do not rupture in the transition. One may confidently assume that this is also the case for the experiments shown in Fig. \ref{Figure1a} and \ref{Figure1b} (left). 2. 
  \item sizes of pores responsible for the permeation process must be of nm-size.
\end{enumerate}
The conduction for ions and small molecules described in this section was named 'macroscopic' permeability because the experiments were made on ensembles of vesicles. This kind of measurement describes the relation of the lipid melting events and the permeability correctly but does not make any statement on the molecular nature of any permeation process. The latter will be discussed in more detail below in section \ref{ChannelsExp}. 

\section{Techniques for measurement of lipid channel activity}\label{Techniques}
The electrical properties of Lipid bilayers and biological membranes are measured through small membrane patches defined by a hole in a hydrophobic substrate \cite{Montal1972} (black lipid membranes), the tip of a glass pipette \cite{Hamill1981} (patch-clamp), or more recently by a hole in a glass surface on a chip \cite{Bruggemann2003} in a technique called 'planar patch-clamp'. Both Montal-Mueller and patch-clamp technique have been used to investigate permeability and channel formation in pure lipid membranes and are briefly described below. 

\subsection{Black lipid membranes / Montal-Mueller technique}
\begin{figure}[htb!]
    \begin{center}
	\includegraphics[width=8cm]{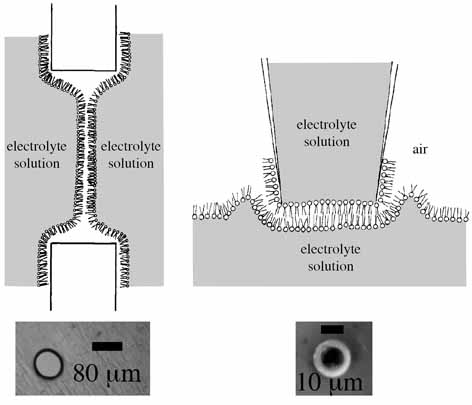}
	\parbox[c]{8cm}
{ \caption{\textit{ Two methods to study currents through artificial membranes. Left: In the Montal-Mueller technique \cite{Montal1972} two monolayers merge into a bilayer over an aperture in a teflon film . Typical aperture size is 50-250 $\mu$m. Right: One can also generate a membrane patch on a glass pipette tip dipped into the surface of a Langmuir monolayer trough \cite{Coronado1983, Hanke1984}. Typical pipette tip sizes are 1-10$\mu$m. The drawings were adapted from \cite{Yoshikawa1988}. The two microscopic images on the bottom show the holes in a teflon film \cite{Fidorra2007} and a pipette tip used for experiments in our lab (courtesy Katrine R. Laub, NBI).}
	\label{Figure2a}}}
    \end{center}
\end{figure}

In 1972, Montal and Mueller \cite{Montal1972} described a method to form lipid bilayers formed over a hole in a thin hydrophobic film (e.g., teflon) from monolayers. Monolayers are formed on an electrolyte surface from a solvent solution, typically containing decane, octane, hexane, pentane, chloroform or other apolar solvents.  The monolayer trough is separated be a teflon film into two compartments. A tiny hole in the film (several 10 $\mu$m)  created by a needle or by an electric spark slightly is located slightly above the monolayer surface. The level of the aqueous buffer is now raised and the monolayers fold over the aperture in the teflon film as shown schematically in Fig. \ref{Figure2a} (left). The monolayers merge into a bilayer at the edge of the teflon aperture. At these edges one expects considerable mechanical tension of quite unclear physics. In order to reduce this tension the teflon aperture is normally pre-painted with hexadecane, a long-chain alkane that is practically water insoluble. Hexadecane must be considered as a impurification that accumulates at the edge of the aperture. It is quite difficult to estimate how much hexadecane is dissolved in the membrane. For this reason, the composition of the membrane is always somewhat unclear even when the composition of the monofilm is exactly known.
Currents are measured with electrodes in the two compartments in a voltage clamp experiment. The capacitance can be measured by continuously changinging voltage such that $dV/dt=const$. The capacitance $C_m$ is obtained from the capacitive current $I_c=C_m\;dV/dt$ necessary to charge the capacitor (assuming that the capacitance itself is a constant, $dC_m/dt=0$, which may not necessarily be correct if voltage changes the state of the membrane). The measurement of the capacitance is required to ensure that only one single membrane has formed. Typical capacitances of membranes are between 0.6 and 1 $\mu$F/cm$^2$.\\
Membranes forming over a hole appear black in visible light due to destructive interference of light reflected from the two bilayer surfaces. They are called black lipid membranes (BLMs) because their formation can be seen by visual inspection under a microscope. The typical hole diameter in a Montal-Mueller experiment is 50 to 250 $\mu$m (Fig. \ref{Figure2a}, left bottom). The disadvantage of this technique is the presence of solvent impurities and the large tension at the edge of the teflon aperture. Furthermore, due to the large diameter of the hole the membranes are often quite unstable. However, this technique is often used to study lipid membrane permeability but also the conductance through channel proteins and peptide pores. 
\subsection{Patch-clamp technique}
The patch-clamp technique has been introduced by Neher and Sakmann \cite{Neher1976} and was described in detail by Hamill et al. \cite{Hamill1981}. It was first used to study channel proteins in biological cells (in particular the acetylcholine receptor in frog muscle cells). A small pipette is placed on a cell surface while applying slight suction. Currents and capacitances are measured by using two electrodes, one being located in the pipette and the other one in the cell. The patch pipettes have much smaller diameters of order 1-10 $\mu$m (see Fig. \ref{Figure2a}, right bottom). The obvious advantage over the Montal-Mueller technique is that one can investigate intact cells and on does not have to use organic solvents. The interface of the patch pipette and the suction is a source for perturbation \cite{Suchyna2009}.\\

Patch pipettes can also be used to study BLMs \cite{Coronado1983, Hanke1984}. The tip of a glass pipette filled with electrolyte solution is placed on a buffer surface. Lipids dissolved in an organic volatile solvent (we use mixtures of hexane and ethanol) are brought into contact with the outer surface of the patch pipette. When the solution runs down the pipette a membrane spontaneously forms at the pipette tip as schematically shown in Fig. \ref{Figure2a} (right). After membrane formation as evident in the capacitance measurement slight suction is applied. In this technique no long chain alkanes are used and one can assume that after some equilibration time the membrane at the pipette tip is practically solvent free (W. Hanke, personal communication). Hanke and coworkers \cite{Hanke1984} also describe a variation of this technique called tip-dipping where the membrane patch is generated from dipping the pipette tip into monolayer surfaces.

\section{Lipid channels}\label{ChannelsExp}

\begin{figure}[htb!]
    \begin{center}
	\includegraphics[width=8.5cm]{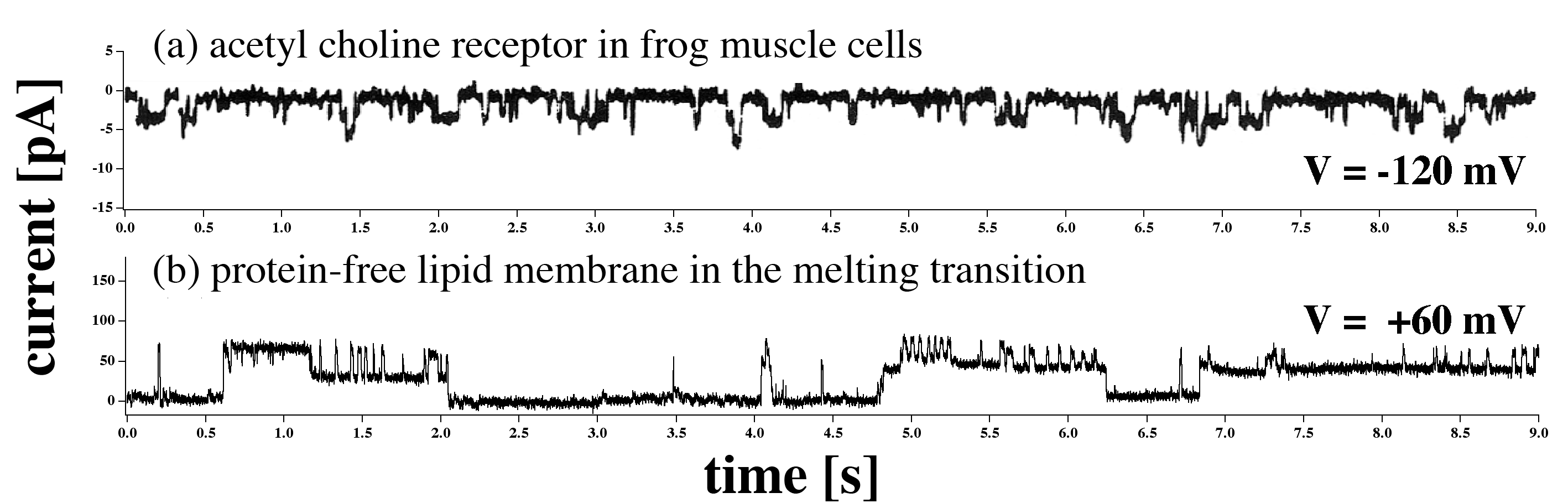}
	\parbox[c]{8cm}
{ \caption{\textit{Comparison of conductance events through biological membranes containing channel proteins and a pure lipid membrane in its melting regime. a. Recording of the acetylcholine receptor in frog muscle cells from the classical paper of Neher \& Sakmann \cite{Neher1976} (Ringer solution in the pipette). b. BLM experiment on a DOPC:DPPC membrane in 150mM KCl, 60mV in its transition regime at 19$^{\circ}$C. Figure adapted from \cite{Heimburg2009}.}
	\label{Figure3_0}}}
    \end{center}
\end{figure}

The evidence for pores (channels) within membrane proteins selectively passing ions rests partially on the observation that the currents through membranes are quantized (Fig.\ref{Figure3_0}a). The fact that the currents are small and correspond to aqueous channels of the dimensions of a few \AA$\,$ suggests that the conduction of ions occurs along objects of molecular dimension. They have been attributed to particular proteins called ion channels. The literature about these proteins is huge (e.g., \cite{Hille1992}). It seems therefore surprising that one can find similar quantized current events in pure lipid membranes that do not contain proteins. Fig. \ref{Figure3_0}a shows the historical experiment of Neher \& Sackmann \cite{Neher1976} on frog muscle cells containing the acetylcholine receptor in the presence of an agonist molecule. Fig. \ref{Figure3_0}b shows the currents through a lipid membrane from a mixture of dioleoyl phosphatidylcholine (DOPC) and dipalmitoyl phosphatidylcholine (DPPC) having a chain melting transition close to the experimental temperature. The melting temperature is important since it has been shown in section \ref{MacroscopicExp} that membranes are most permeable in their melting transition. It is obvious that the two traces display very similar features both regarding amplitude and time scales of the currents. In the following we will characterize the lipid channels in more detail.\\
\subsection{Early experiments}\label{ChannelsExp.1}
\begin{figure}[htb!]
    \begin{center}
	\includegraphics[width=6.5cm]{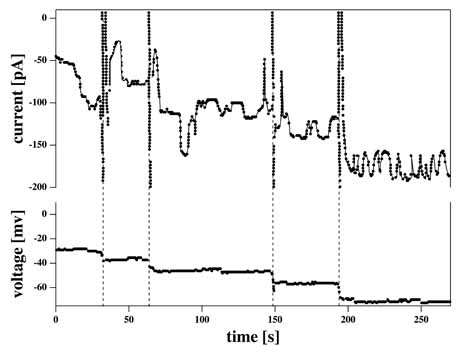}\\
	\includegraphics[width=6.5cm]{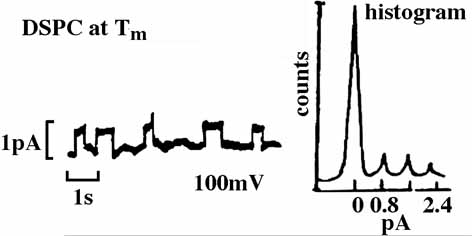}	
	\parbox[c]{8cm}
{ \caption{\textit{Top: First channel recordings in a lipid membrane made of oxidized cholesterol \cite{Yafuso1974}. The bottom line indicates the voltage. The dashed lines indicate the voltage changes that give rise to large capacitive currents in the current recording. For constant voltage current fluctuations of the order of 10-20 pA occur. The data were digitally adapted from the original Figure and numerically corrected for the distortions caused by the analogue chart-recorder. Bottom: Quantized currents through synthetic DSPC membranes close to their transition in 1M KCl at 100mV \cite{Antonov1980}. }
	\label{Figure3ab}}}
    \end{center}
\end{figure}

\begin{figure}[htb!]
    \begin{center}
	\includegraphics[width=8cm]{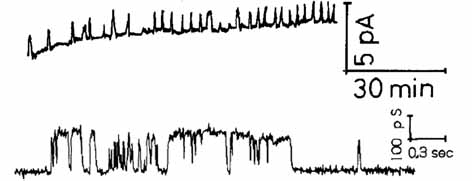}
	\parbox[c]{8cm}
{ \caption{\textit{Top: Current fluctuations in a mixture of DOPC and cholesterol (about 9:1) at 20mV in 1 M  KCl \cite{Yoshikawa1988}. Bottom: Conductance fluctuations of a membrane made of a POPE:POPC=7:3 mixture in 150mM KCl \cite{Woodbury1989}.}
	\label{Figure3c}}}
    \end{center}
\end{figure}

To the best of our knowledge the first channel events recorded in pure lipid membranes are from Yafuso et al. in 1974 \cite{Yafuso1974}, two years before the famous paper by Neher and Sakmann on the acetylcholine receptor shown in Fig. \ref{Figure3_0}a. Their experiments on bilayers from oxidized cholesterol are shown in Fig. \ref{Figure3ab} (left). 
They found spontaneous conductance changes and multilevel conductance states in an unmodified synthetic lipid film (of somewhat exotic nature). While the lipid system used does not allow to draw conclusion with respect to biological membranes, the experiments quite clearly are a proof-of-principle in so-far as they demonstrate that lipid films alone can generate phenomena that are typically attribute to proteins, peptide pores or other additives.  The original recording was made with an analogue chart recorder in which the pen position was changed by rotation around an axis. This has been digitally corrected by us (see the original paper for a reference to the raw data). A can be seen in  Fig. \ref{Figure3ab} (top) a change of the voltage leads to an overall larger membrane current the shows quantized steps.\\ 
Another important early paper showing this phenomenon is authored by Antonov and collaborators in 1980 \cite{Antonov1980}. They showed that one obtains quantized currents DSPC membranes one obtains quantized currents of in the pA regime at 100 mV. As in the paper by Yafuso et al. \cite{Yafuso1974}, they found that current traces may display several steps as shown in the current histogram in Fig.  \ref{Figure3ab} (bottom). In their paper they state: "We suggest that these channels could conduct the transmembrane ionic current in biological membranes without the involvement of peptides and proteins", thus making the first statement indicating that this could be a general phenomenon in biomembranes.\\
Other authors have found the same phenomenon. Fig. \ref{Figure3c} (top) shows results from Yoshikawa et al. \cite{Yoshikawa1988}. They described quantized current fluctuations through DOPC membranes containing some cholesterol that are of order 1 pA (corresponding to a conductance of 50 pS). The authors found similar results for membranes in BLMs, on patch pipettes and on membranes deposited on filter paper. As other authors, they also discuss the relevance of this finding for the interpretation of currents in biomembranes and suggest that currents attributed to proteins can also arise from lipid membrane fluctuations.\\
\begin{figure}[htb!]
    \begin{center}
	\includegraphics[width=8cm]{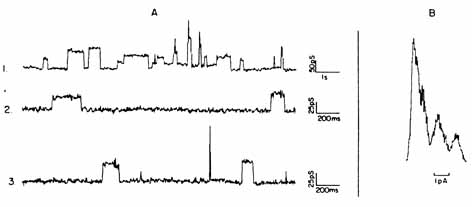}
	\parbox[c]{8cm}
{ \caption{\textit{Currents through a purely synthetic diphytanoyl phosphatidylcholine membrane (1M KCl unbuffered, pH ca. 6.5). Left: Currents recorded at 77mV in different scalings. Right: Current histogram showing several steps. From Kaufmann et al. \cite{Kaufmann1989c}.}
	\label{Figure3c2}}}
    \end{center}
\end{figure}
Fig. \ref{Figure3c} (bottom) shows quantized membrane currents in a membrane made from POPE:POPC=7:3 mixture (from \cite{Woodbury1989}). The currents display a similar conductance as in the above experiment.
The authors discuss the similarity of the lipid conduction steps with those attributed to proteins channels and state: "This observation demonstrates the need for caution in interpreting conductance changes, which occur following ejection of channel- containing vesicles near a membrane." This refers to the standard reconstitution procedure of generating synthetic membranes containing proteins. In particular, the authors state that currents attributed to proteins could in fact be due to lipid pores.\\
Kaufmann et al. \cite{Kaufmann1989c} showed further interesting results on channel appearance in purely synthetic membranes. Their paper appeared as a booklet (available from the website given in the reference list and from the University Library in G\"ottingen, Germany) and gives considerable insight into the physics of the pore formation process that goes along the line of argument in section \ref{Theory.2}. 
\begin{figure}[htb!]
    \begin{center}
	\includegraphics[width=6cm]{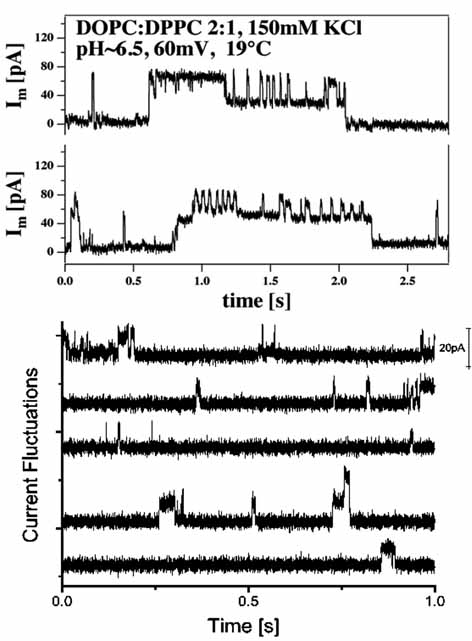}
	\parbox[c]{8cm}
{ \caption{\textit{Top: Current fluctuations in a DOPC:DPPC=2:1 mixture at 19$^{\circ}$C (150mM KCl, 60mV). One can resolve at least 4 quantized steps with a current amplitude of about 20 pA corresponding to about to a conductance of about 300pS. From \cite{Blicher2009}. Bottom: Current fluctuations in a  DC$_{15}$PC:DOPC=95:5 mixture at 31.5$^{\circ}$C (1000mV), which is the transition regime of this mixture. Quantized currents with an amplitude of 15pA can be seen. From \cite{Wunderlich2009}.}
	\label{Figure3d}}}
    \end{center}
\end{figure}
A nice example for quantized currents through membranes is given in Fig. \ref{Figure3d} (top) \cite{Blicher2009}. At least 4 steps in the current recording through a DOPC:DPPC= 2:1 mixture can be seen. Each of them corresponds to a conductance of about 300 pS. The experimental temperature of 19$^{\circ}$C is close to the transition maximum of this lipid mixture.  Fig. \ref{Figure3d} (bottom) shows a long trace from a BLM made of a DC$_{15}$PC:DOPC=95:5  mixture in the phase transition regime   (31.5$^{\circ}$C, 1000mV) \cite{Wunderlich2009}. The above experiments show that there is quite a range of possible pore sizes and conductances found for lipid membranes.\\
We focus here on literature demonstrating quantized current events through lipid membranes. There is additionally a complete literature on reversible electrical breakdowns of artificial membrane conductance at higher voltages \cite{Benz1979, Abidor1982, Neumann1999}, which are connected to electroporation. We believe, that lipid channel formation and electroporation are related phenomena.

\subsection{Dependence of channel formation of the melting transition}\label{ChannelsExp.2}
As mentioned in section \ref{MacroscopicExp} the overall membrane conductivity is maximum in the transition regime. Consistent with this finding the likelihood of finding lipid channel events is maximum in the lipid melting regime. Below we show three experiments that demonstrate this. 
\begin{figure}[b!]
    \begin{center}
	\includegraphics[width=8cm]{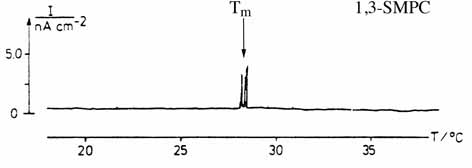}	
	\parbox[c]{8cm}
{ \caption{\textit{Currents through a 1,3-SMPC membrane with a transition at 29$^{\circ}$ recording during a temperature scan. At the transition temperature quantized current events were found (not well resolved in the figure but well described in the text). From \cite{Boheim1980}. }
	\label{Figure3b2}}}
    \end{center}
\end{figure}
Boheim et al. \cite{Boheim1980} reported quantized current events in synthetic 1,3-SMPC membrane at the phase transition temperature of about 29$^{\circ}$C (Fig. \ref{Figure3b2}). The discrete conductance steps are not well visible in the figure, but were obvious in the original data trace (Wolfgang Hanke, personal communication). They write: "Apparently, the transition of a bilayer membrane from the liquid-crystalline to the solid state and vice versa is associated with spontaneous fluctuations in membrane conductance."
\begin{figure}[htb!]
    \begin{center}
	\includegraphics[width=8cm]{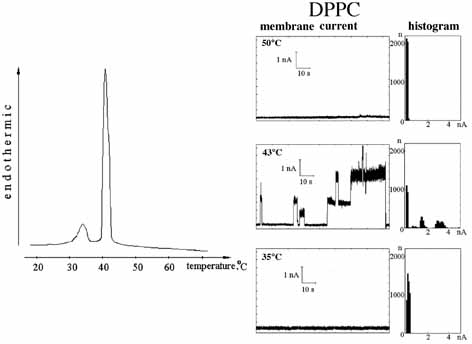}
	\parbox[c]{8cm}
{ \caption{\textit{Currents through DPPC membranes with $T_m$ around 41$^{\circ}$C. Left: heat capacity profile of DPPC. Right: Currents through DPPC membranes above (50$^{\circ}$C), close to (43$^{\circ}$C) and below (35$^{\circ}$C) the transition. Only close to the transition quantized currents with discrete steps in the current histogram were seen. From \cite{Antonov2005}.}
	\label{Figure3f}}}
    \end{center}
\end{figure}
Antonov et al. demonstrated for both DSPC \cite{Antonov1980} and DPPC \cite{Antonov2005} that they found quantized currents in the phase transition regime while the membrane seemed inert outside of this regime. Fig.\ref{Figure3f} (left) shows DPPC membranes above the transition at 50$^{\circ}$C (top), in the melting regime at 43$^{\circ}$C and below the transition at 35$^{\circ}$C. The current fluctuations in the transition are quantized and are of unusual magnitude (nA regime indicating very large pores).\\ 
Transitions in single lipid membranes occur over a quite narrow temperature interval of less than one degree which makes it difficult to adjust the experimental temperature accurately. The heat capacity displays a very large amplitude at maximum. As shown below (section \ref{Theory.5}) that the permeability is closely related to the heat capacity, and that fluctuations in area are very large. Thus, in transitions of single lipid membranes they display a tendency to show very large pores and a pronounced tendency to rupture. Those black lipid membranes are inherently unstable and break easily.\\
For the above reason, other authors prefer to work with lipid mixtures. They display broader melting profiles with smaller fluctuations at maximum. These membranes are more stable and due to their wider melting regime not as temperature sensitive. This approach has been taken by \cite{Blicher2009, Fidorra2007}. They used a DOPC:DPPC=3:1 mixture (150mM KCl, 40mV) that displays a broad melting profile with a maximum around 17$^{\circ}$C.At the heat capacity maximum many discrete current steps of about 30pA were found (Fig. \ref{Figure3g}), corresponding to a conductance of $\approx$ 750 pS. Above the melting temperature at 30 $^{\circ}$C (in the fluid phase regime) no discrete current steps were seen. 
\begin{figure}[htb!]
    \begin{center}
	\includegraphics[width=8cm]{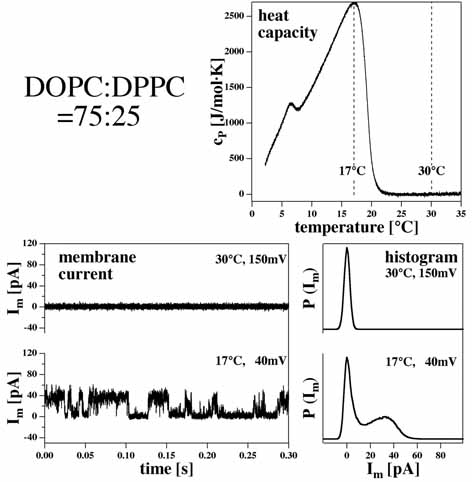}
	\parbox[c]{8cm}
{ \caption{\textit{Currents through BLMs of a DOPC:DPPC=3:1 membrane (from \cite{Blicher2009}). Top: heat capacity profile with a maximum at 17$^{\circ}$C. The two dotted lines represent the temperatures at which the currents in the bottom panel were recorded. Bottom, left: Current steps could only be observed at the transition temperature (17$^{\circ}$C, 150mM KCl, 40mV) but not in the fluid phase regime at 30$^{\circ}$C. Bottom, right: Current histograms. A baseline current was subtracted.}
	\label{Figure3g}}}
    \end{center}
\end{figure}

\subsection{Specificity of lipid channels}\label{ChannelsExp.3}
In their seminal paper from 1980, Antonov et al. \cite{Antonov1980} compared quantized currents through membranes made of  the zwitterionic DSPC with those through positively charged membranes of a DSPC analogue in which the P-O$^-$ in the head-group was replaced by a P-CH$_3$ rendering the membrane positively charged. They found that the cationic membranes were anion selective while zwitterionic membranes did not display any selectivity for charges (unfortunately, no experimental details were given).\\
\begin{figure}[htb!]
    \begin{center}
	\includegraphics[width=6cm]{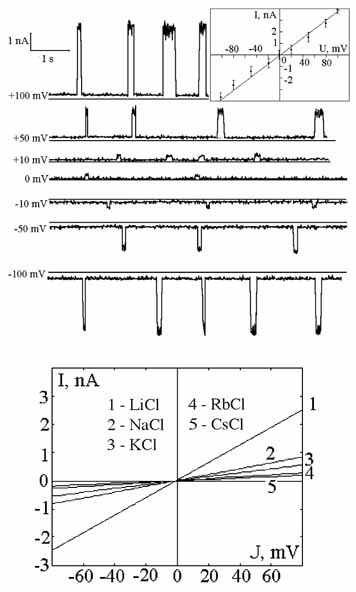}
	\parbox[c]{8cm}
{ \caption{\textit{Currents through pores in DPPC membranes close to the chain melting temperature (43$^{\circ}$). Top: Current traces in the presence of LiCl at different voltages lead to a linear current-voltage relationship. Bottom: Current-voltage relationships in the presence of LiCl, NaCl, KCl, RbCl and CsCl yield different conductances. }
	\label{Figure3e}}}
    \end{center}
\end{figure}
In \cite{Antonov2005}, the same group of authors compared the conductances of membrane pores for different cations. They recorded the single channel currents through DPPC membrane in the presence of LiCl, NaCl, KCl, RbCl and CsCl. In a voltage regime from -80 to +80 mV they found a linear current-voltage relationship from which a conductance $g_i$ can be calculated ($I_i=g_i\cdot U$, where  $i$ is the index for the ion species). The details of the experiment are given in Fig. \ref{Figure3e}. The authors found that the value of the conductance decreases in the sequence:
\begin{displaymath}
g_{Li} > g_{Na} > g_K > g_{Rb} > g_{Cs}
\end{displaymath}
Antonov and collaborators discussed this sequence in connection with kosmotropes (water structure enhancers) and chaotropes (water structure breakers) that are closely related to the Hofmeister series. This series involves a ranking of ions in respect to their effect on various materials in electrolytes (e.g., proteins) and is related to the interaction of the ions with water. The Hofmeister series has been discussed in connection with peptide pore conductances \cite{Gurnev2009}. According to \cite{Cacace1997} the sequence from chaotropic to kosmotropic  for the above ions is $Li^+  >  Cs^+  >  Na^+  >  K^+$ while it was given as $Na^+ > K^+  >  Rb^+  > CS^+$ by M. Chaplin\footnote{See: www1.lsbu.ac.uk/water/hofmeist.html} consistent with the conductances in the above experiment. It is very likely that such correlations exists, rendering the lipid membrane moderately ion selective\footnote{Selectivity is considered here to be a conduction preference expressed in different permeation rates rather than an exclusive all-or-nothing conduction phenomenon.}.\\

\section{Theoretical considerations}\label{Theory}

\subsection{Pore models}\label{Theory.1}

The remarkable finding in electrophysiological lipid membrane experiments is that ion permeation expresses itself as quantized current events. This all-or-nothing feature is unexpected since one might expect the membrane defects form through stochastic processes that favor broad distributions of pore sizes. 
\begin{figure}[htb!]
    \begin{center}
	\includegraphics[width=4cm]{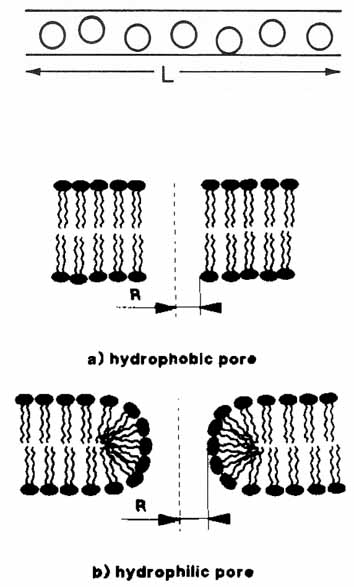}
	\parbox[c]{8cm}
{ \caption{\textit{Two representations for transport across a membrane. Top: Single file transport. Water molecules align in a pore. From Finkelstein \cite{Finkelstein1987}. Center: Schematic drawing of membrane pores:  a hydrophobic pore. Bottom: a hydrophilic pore. From Glaser et al. \cite{Glaser1988}.}
	\label{Figure5a}}}
    \end{center}
\end{figure}
The permeability of membranes has been imagined to occur via three mechanisms:
\begin{enumerate}
  \item diffusion of solutes through the hydrophobic core of the membrane
  \item transport along water files consisting of single H$_2$O molecules (Fig.\ref{Figure5a},left)
  \item hydrophobic pore formation (Fig.\ref{Figure5a}, right top)
  \item hydrophilic pore formation (Fig.\ref{Figure5a}, right bottom)
\end{enumerate}
Mechanism 1 was proposed for oil-soluble substances like anesthetics already in the 19$^{th}$ century \cite{Overton1895}. However, for ions or polar molecules as water this is an unlikely mechanism considering the very low solubility of such substance in the hydrophobic core membrane core. The free energy of a charge in a dielectric medium is proportional to the electrostatic potential $\Psi=q/\epsilon_0 \epsilon r$ where q is the charge, $\epsilon_0$ is the vacuum permittivity, $\epsilon$ is the dielectric constant, and $r$ the distance from the charge. The dielectric constant in water is about 80 while it is 2-4 in the membrane interior. Thus, the electrostatic free energy is different by about a factor of 20-40 between membranes and water. Due to Debye-H\"uckel ion screening this effect is even larger in the aqueous electrolyte solutions. \\
\begin{figure}[htb!]
    \begin{center}
	\includegraphics[width=7cm]{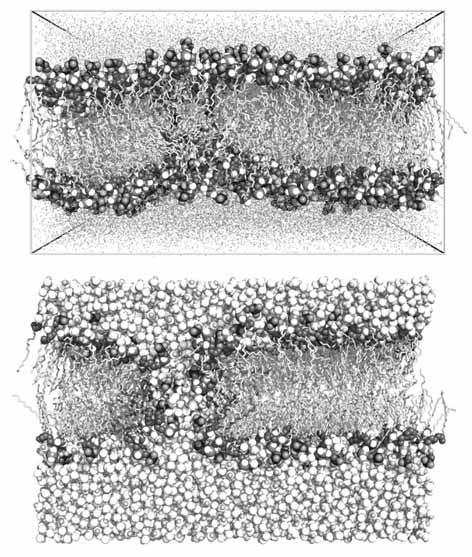}
	\parbox[c]{8cm}
{ \caption{\textit{Molecular dynamics simulation of a POPC membrane with a tansmembrane voltage of 2V. Top: Hydrophobic prepore in an early stage of the simulation (less than 1ns). Bottom: Pore with dimensions of the order of 1 nm after about 50ns. Adapted from B\"ockmann et al. \cite{Boeckmann2008}.}
	\label{Figure5b}}}
    \end{center}
\end{figure}
Transport mechanism 2 has been described by Levitt \cite{Levitt1984} and Finkelstein \cite{Finkelstein1987}, see Fig. \ref{Figure5a} (top). Finkelstein defines: "By single-file transport of water one means that water molecules within the pore cannot pass or overtake each other." In this context the term "no-pass transport" has been used.  Ions would be part of the file and can be transported in sequence with the water molecules. The single file mechanism might explain the quantized nature of currents because there is a minimum size for the water file defined by the dimensions of H$_2$O and the ions. \\
Pore formation is the most popular mechanism proposed for membrane transport. Glaser et al. \cite{Glaser1988} suggested both hydrophobic (mechanism 3, Fig.\ref{Figure5a}, center) and hydrophilic pores (mechanism 4, Fig.\ref{Figure5a}, bottom). The hydrophobic pore is probably related to the formation of a water file as described in (2) and does not involve any rearrangement of lipids. The hydrophilic pore involves a change in the geometry of the lipid head groups. The elastic free energy of such a pore can be calculated assuming typical values for the elastic constants. Stable pores with a diameter of $\sim$ 1 nm were suggested. On theoretical grounds it was further proposed that transmembrane voltage can stabilize large pores at high voltage \cite{Winterhalter1987}. This might be important in connection with the phenomenon of electroporation \cite{Neumann1999}.\\
\begin{figure}[htb!]
    \begin{center}
	\includegraphics[width=8cm]{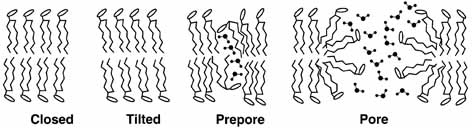}
	\parbox[c]{8cm}
{ \caption{\textit{Schematic drawing of the chain of events in the molecular dynamics simulation of POPC membranes shown in Fig. \ref{Figure5b}. After the application of a voltage of 2V across the membrane, first a "prepore" opened that is both consistent with the water file shown in Fig. \ref{Figure5a} (top) and the hydrophobic pore in Fig. \ref{Figure5a} (center). Subsequently, a hydrophilic pore with a radius of about 0.5 nm formed. From B\"ockmann et al. \cite{Boeckmann2008}.}
	\label{Figure5c}}}
    \end{center}
\end{figure}
Recent molecular dynamics simulations of membranes exposed to large voltages ($\sim$ 2V) suggest that both single file and hydrophilic pore formation can occur. Fig. \ref{Figure5b} shows a POPC membrane at two stages during the simulation after a voltage of $\sim 2V$ was applied. In Fig. \ref{Figure5b} (top) a single water file formed along which ions could flow. This was called a 'prepore' \cite{Boeckmann2008}  and occurred after less than 1 ns of simulation time.  After $\sim$ 50ns a large pore developed long this defect  that had a diameter about 1 nm (Fig. \ref{Figure5b}, bottom). The chain of  events in the simulation is shown in Fig. \ref{Figure5c}. First, a water file (prepore) forms along which ions can flow. This is consistent with the hydrophobic pore suggested by \cite{Glaser1988} (Fig. \ref{Figure5a}, center). Subsequently, the lipids rearrange and form a hydrophilic pore  (Fig. \ref{Figure5a}, bottom).

If we assume that this is the likely chain of events then we would conclude that the single file (prepore) defines a minimum size of a conducting event in a membrane. The hydrophilic pore seems to display a stable radius of the order of 0.5 nm. This is consistent with the magnitude of the currents measured in the patch-clamp and the Montal-Mueller techniques. Remarkably, this is also the typical size of the aqueous pores proposed for channel proteins. \\
Assuming a pore of radius r, Glaser et al. \cite{Glaser1988} calculated a free energy of pore formation of
\begin{equation}
\label{eq_Theory.1}
\Delta F(r)=2\pi \gamma r+ \pi r^2 (\sigma + \alpha U^2)
\end{equation} 
where $\gamma$ is the edge tension of the pore, $\sigma$ the lateral tension and $U$ the voltage. $\alpha$ is a coefficient related to the capacitance of the membrane. The free energy therefore consists of three contributions: the energy of the pore interface, the work that has to be performed against an external tension $\sigma$ that is proportional to the pore area, and a correction term containing voltage that is also proportional to the pore area. One should assume that the probability of pore formation is related to $\exp(-\Delta F(r)/kT$) and the conductance of one pore is in a simple manner related to the pore radius $r$.\\

The above equation does not yet contain a particular theory for the effect of the phase transition, which is obviously of major importance as judged by the experimental data shown in sections \ref{MacroscopicExp} and \ref{ChannelsExp}. This will be provided below. 

\subsection{Entropy, thermodynamic variables and forces}\label{Theory.2}
In 1910, Albert Einstein wrote a remarkable paper on critical opalescence \cite{Einstein1910} in which he considered the phenomenon of maximum light scattering of liquid mixing close to a critical point (called ''critical opalescence"). 
This paper contains a deep reflection of the nature of the second law of thermodynamics and is at the basis of all non-equilibrium thermodynamics and fluctuation theory. The second law basically states that the most likely state of a system is most probable. This were without content if one could not extract more information from this law without making assumptions on the molecular detail of a system. Einstein showed in fact that one can make general statements about the nature of fluctuations. His theory for critical opalescence is similar to the thermodynamic theory of the pore formation process outlined below. The application of this theory for pore formation was already proposed by Nagle \& Scott \cite{Nagle1978b} and Kaufmann et al. \cite{Kaufmann1989c}.\\

Let us consider the entropy as a function of the extensive thermodynamic variables $\alpha$ (e.g., energy E, volume V, area A, charge q, ...) and define the deviations from the equilibrium value $\xi_i=\alpha_i-\alpha_{i,0}$ we can expand it into a Taylor series:
\begin{equation}
\label{eq_Theory.2.1}
S=S_0+\sum_i\left(\frac{\partial S}{\partial \xi_i}\right)_0 \xi_i+\frac{1}{2}\sum_{ij}\left(\frac{\partial^2 S}{\partial \xi_i\xi_j}\right)_0 \xi_i \xi_j+ ...
\end{equation}
The fact that the entropy is at maximum in an equilibrated system implies that the linear terms $(\partial S/\partial \xi_i)_0$ are zero and thus we can approximate the entropy for small fluctuations with 
\begin{equation}
\label{eq_Theory.2.2}
S\approx S_0+\frac{1}{2}\sum_{ij}\left(\frac{\partial^2 S}{\partial \xi_i\xi_j}\right)_0 \xi_i \xi_j \equiv S_0-\sum_{ij} g_{ij} \xi_i \xi_j
\end{equation}
where $g_{ij}=-0.5 (\partial^2 S /\partial \xi_i\xi_j)_0$ are positive constants.\\
\begin{figure}[htb!]
    \begin{center}
	\includegraphics[width=6cm]{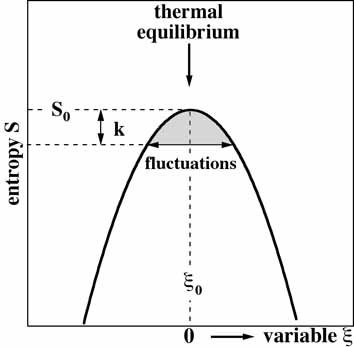}
	\parbox[c]{8cm}
{ \caption{\textit{The entropy as a harmonic potential and the corresponding fluctuations of the variable $\xi$ that could be internal energy, volume, area etc. The slope of the potential at $\xi$ is the thermodynamic force, while the curvature of the potential at $\xi_0$ is proportional to the corresponding susceptibility (cf. eqs. \ref{eq_Theory.2.6}, \ref{eq_Theory.2.9} and \ref{eq_Theory.3.2}).}
	\label{Figure5d}}}
    \end{center}
\end{figure}

The expansion of the entropy into a quadratic function of the intensive thermodynamic variables is at the basis of the complete discipline of linear non-equilibrium thermodynamics as developed by Onsager \cite{Onsager1931a, Onsager1931b} and Prigogine \cite{Kondepudi1998}.
The probability to find a particular pair of fluctuations, $\xi_i$ and $\xi_j$,  is given by
\begin{equation}
\label{eq_Theory.2.3}
P(\xi_i, \xi_j)=P_0\exp\left(\frac{S(\xi_i,\xi_j)}{k}\right)=P_0\exp\left(-\frac{g_{ij} \xi_i \xi_j}{k}\right)
\end{equation}
or for one single fluctuation $\xi_i$  it is given by
\begin{equation}
\label{eq_Theory.2.4}
P(\xi_i)=P_0\exp\left(-\frac{g_{ii} \xi_i^2}{k}\right)
\end{equation}
which corresponds to a Gaussian distribution of states with $P_0=\sqrt{g_{ii}/\pi\;k}$.\\
The thermodynamic average of the product of the fluctuations $\xi_i\xi_j$ is given by \cite{Greene1951}
\begin{equation}
\label{eq_Theory.2.5}
 \left<\xi_i \xi_j\right>=\int \xi_i \xi_j P(\xi_i, \xi_j) d\xi_1 d\xi_2 d\xi_3 ...d\xi_n =-\frac{k}{2g_{ij}} 
\end{equation}
where $\left<...\right>$  denotes the statistical mean, which implies
\begin{equation}
\label{eq_Theory.2.6}
\left<\xi_i^2\right>=-\frac{k}{2g_{ii}} 
\end{equation}
This corresponds to the mean square amplitude of the fluctuations or the width of the Gaussian distribution. This implies that the half width of the distribution is related to the curvature of the entropy potential, a fact that we will make use of below in order to derive the relaxation times of fluctuations.\\

The thermodynamic forces are defined as the derivatives of the entropy  with respect to the extensive variable.
\begin{equation}
\label{eq_Theory.2.7}
X_i=\frac{\partial S}{\partial \xi_i}=-\sum_j g_{ij}\xi_j
\end{equation}
This is completely in analogy to the definition of forces in classical mechanics (as the derivative of a potential) but more general since it includes extensive variable as the energy and charge, but also the particle numbers . Further, the potential under consideration here is the entropy.

Some of the thermodynamics forces are
\begin{eqnarray}
\label{eq_Theory.2.8}
\frac{1}{T} & \quad\mbox{with the conjugated variable}\quad & E \quad\mbox{(interal energy)} \nonumber\\
\frac{p}{T} & \quad\mbox{with the conjugated variable}\quad &V \quad\mbox{(volume)}\nonumber\\ 
\frac{\Pi}{T} & \quad\mbox{with the conjugated variable}\quad & A\quad\mbox{(area)} \\
\frac{\Psi}{T} & \quad\mbox{with the conjugated variable}\quad & q\quad\mbox{(charge)} \nonumber\\
\frac{\mu_j}{T} & \quad\mbox{with the conjugated variable}\quad & n_j\quad\mbox{(particle number)} \nonumber\\
 &...& \nonumber
\end{eqnarray}
where $T$ is the temperature, $p$ is the pressure, $\Pi$ is the lateral pressure, $\Psi$ is the electrostatic potential, and $\mu_j$ is the chemical potential of species $j$ \cite{Greene1951, Kondepudi1998}. E, V, A, q and n$_j$ correspond to the fluctuating variables $\xi_i$. 

The above implies that
\begin{equation}
\label{eq_Theory.2.9}
\left<\xi_i^2\right>=-k\left(\frac{\partial X_i}{\partial \xi_i}\right)^{-1}=kg_{ii}^{-1}
\end{equation}
meaning that the fluctuations are related both to the curvature of the entropy potential of the change of the forces upon the change of a variable. Eq. \ref{eq_Theory.2.9} has been strictly derived in \cite{Greene1951}.

\subsection{Susceptibilities}\label{Theory.3}
The susceptibilities or response functions are functions like the heat capacity, the compressibility or the capacitance. They are given by the derivative of an extensive variable with respect to an intensive variable. E.g.,
the heat capacity at constant volume is given by
\begin{equation}
\label{eq_Theory.3.1}
c_V=\left(\frac{\partial Q}{\partial T}\right)_V=\left(\frac{\partial E}{\partial T}\right)_V
\end{equation}
Now, $\xi_E\equiv E-\left<E\right>$ is one of the quantities that can fluctuate, while $\partial T=-T^2\partial (1/T)$ is related to the conjugated thermodynamic force, $X_E=1/T$. In particular,
\begin{equation}
\label{eq_Theory.3.2}
c_V=-\frac{1}{T^2}\left(\frac{\partial \xi_E}{\partial X_E}\right)_V=-\frac{1}{T^2}\left(\frac{\partial X_E}{\partial \xi_E}\right)_V^{-1}\stackrel{eq.\ref{eq_Theory.2.9}}{=}+\frac{\left<\xi_E^2\right>}{kT^2}
\end{equation}
Another way of writing this is
\begin{equation}
\label{eq_Theory.3.3}
c_V=\frac{\left<E^2\right>-\left<E\right>^2}{kT^2}
\end{equation}
This implies that the heat capacity is proportional to the fluctuations in internal energy. This relationship could also be derived in a simpler manner by just differentiating the statistical mean of the internal energy that is given by $\left<E\right>=\sum_i E_i \exp(-E_i/kT)/\sum_i \exp(-E_i/kT)$. Thus, the fluctuation relations would be strictly true even if the assumption of a harmonic entropy potential is not made. Similar relations can be written for the heat capacity at constant pressure
\begin{equation}
\label{eq_Theory.3.4}
c_p=\frac{\left<H^2\right>-\left<H\right>^2}{kT^2}
\end{equation}
the isothermal volume compressibility
\begin{equation}
\label{eq_Theory.3.5}
\kappa_T^V=-\frac{1}{V}\left(\frac{\partial V}{\partial p}\right)_T=\frac{\left<V^2\right>-\left<V\right>^2}{V \;kT}
\end{equation}
and the isothermal area compressibility
\begin{equation}
\label{eq_Theory.3.6}
\kappa_T^A=-\frac{1}{A}\left(\frac{\partial A}{\partial \Pi}\right)_T=\frac{\left<A^2\right>-\left<A\right>^2}{A\;kT}
\end{equation}
Eqs. \ref{eq_Theory.3.3} to \ref{eq_Theory.3.6} are linked to the fluctuation theory (see, e.g., \cite{Greene1951}). The principal underlying idea is that small forces generated thermal motion generate local fluctuations in the extensive variables. The coupling of the susceptibilities to the fluctuations is important and will play an crucial role below.

\subsection{Coupling of compressibilities with the heat capacity}\label{Theory.4}
It has been argued by \cite{Heimburg1998, Ebel2001} that for transitions in membranes one finds that
\begin{equation}
\label{eq_Theory.4.1}
\Delta V(T)=\gamma_V\cdot \Delta H(T)
\end{equation}
where $\Delta V(T)$ is the temperature dependent volume change in the transition regime, and $\Delta H(T)$ is the corresponding change in enthalpy. This means that volume and enthalpy change in a proportional manner with a proportional constant $\gamma_V=7.8 \cdot 10^{-10}$m$^3$/J for many lipid systems including biological membranes \cite{Ebel2001}.
More indirectly, it has been argued that a similar relation is also true for area changes in melting transitions of membranes:
\begin{equation}
\label{eq_Theory.4.2}
\Delta A(T)=\gamma_A\cdot \Delta H(T)
\end{equation}
with $\gamma_A=0.89$m$^2$/J \cite{Heimburg1998}. The two above relation together with eqs. \ref{eq_Theory.3.5} and \ref {eq_Theory.3.6} imply that the fluctuations of enthalpy are proportional to the fluctuations in volume or area (i.e., enthalpy, volume and area are no independent variables), and we obtain:
\begin{equation}
\label{eq_Theory.4.3}
\Delta\kappa_T^V=\frac{\gamma_V^2T}{V}\Delta c_p\qquad\mbox{and}\qquad\Delta\kappa_T^A=\frac{\gamma_A^2T}{A}\Delta c_p
\end{equation}
This means that both volume and area compressibility are proportional to the heat capacity changes in membrane transitions. In particular it implies that membranes are very soft in their transitions. This change in compressibility can easily of order 10-fold at the transition maximum. This has profound consequences for the pore formation process.

\subsection{Permeability}\label{Theory.5}
A couple of authors have discussed area fluctuations as the origin of membrane pores (or lipid ion channels), e.g. \cite{Papahadjopoulos1973, Cruzeiro1988, Jansen1995, Blicher2009, Wodzinska2009} but in particular Nagle \& Scott \cite{Nagle1978b} and Kaufmann \cite{Kaufmann1989c} (much of the line of argument in sections \ref{Theory.2} and \ref{Theory.3} was given in the latter reference and applied to membrane pores).\\
There are in particular two views:
\begin{enumerate}
  \item In the transition one finds phase coexistence and the likelihood of finding a membrane pore is at maximum at the domains boundary. Therefore, the permeability is proportional to the length of the domain interfaces. This view has been taken by Papahadjpoulos et al. \cite{Papahadjopoulos1973} and Mouritsen and coworkers \cite{Cruzeiro1988, Corvera1992}.
  \item In order for a membrane pore to be created work has to be performed that is proportional to the lateral compressibility \cite{Nagle1978b, Blicher2009, Wodzinska2009}.  
\end{enumerate}
These two views are conceptually similar but not identical. Clearly, the domains occur in temperature regimes with larger area fluctuations. More precisely, the fluctuations in domain sizes are responsible for both the large heat capacity and the large compressibility. Domain sizes in lipid transitions depend on the cooperativity of the melting process. In computer simulations, cooperativity is usually modeled by an unlike nearest neighbor interaction that is effectively a free energy penalty for the creation of a domain interface. For this reason, more cooperative transitions with narrower and more pronounced heat capacity peaks display larger domains at the melting temperature (50\% of the lipid matrix is in a liquid state) and consequently a smaller overall domain interface than less cooperative transitions. 
Thus, large fluctuations imply that domains are large and that the total domain interface is smaller. The data shown, e.g., in Fig. \ref{Figure1b} (right) indicate, however, that the permeability is about proportional to the heat capacity. This favors the second view. The discrepancy between the two views would disappear if one would not consider domain interface is one-dimensional lines but rather as narrow areas of large fluctuations. However, under these circumstances, one would have to take both length and thickness of the interfacial region into account. For these reasons we follow the argument that the overall fluctuations are responsible for the increase in permeability. Domain interfaces are discussed again in section \ref{Interfaces1} and it is shown that the fluctuations within different domains may also be different.\\

Nagle and Scott \cite{Nagle1978b} calculated the free energy (the work) necessary for the formation of a pore of area $\Delta A$ in a membrane with total area to be
\begin{equation}
\label{eq_Theory.5.1}
\Delta F=\frac{1}{2}\frac{1}{\kappa_T^A A}(\Delta A)^2
\end{equation}
with the lateral compressibility $\kappa_T^A$  of the membrane. This corresponds to Hooke's law for a mechanical spring.
The likelihood to find a pore of size $\Delta A$ is given by
\begin{equation}
\label{eq_Theory.5.2}
P(\Delta A)=P_0\exp \left(-\frac{\Delta F}{kT}\right)
\end{equation}
which implies that the likelihood for a pore is the same when $(\Delta A)^2/\kappa_T^A$ is the same. This means one finds pores of twice the area if the compressibility is four times larger, or that the likely pore size scales with
\begin{equation}
\label{eq_Theory.5.3}
(\Delta A)^2\propto \kappa_T^A
\end{equation}
As a next step it was assumed that the permeability $P$ of the membrane is related to the size of the defects such that one can make a series expansion of the permeability as a function of area change $\Delta A$:
\begin{equation}
\label{eq_Theory.5.4}
P=P_0+C_1\Delta A +C_2(\Delta A)^2+....
\end{equation}
where $P_0$ is the permeability of the pure phase in the absence of fluctuations. In the transition one expects fluctuations $\Delta A$ that are both positive and negative and the linear term is neglected, yielding
\begin{equation}
\label{eq_Theory.5.5}
P=P_0+C_2(\Delta A)^2\stackrel{eq.\ref{eq_Theory.5.3}}{=}P_0+C_2' \kappa_T^A
\end{equation}
This equation reflects that one expects larger pore areas $\Delta A$ or a larger number of pores if the compressibility is higher. Nagle \& Scott \cite{Nagle1978b} argued that one obtains the same law if one takes into account that one finds distributions of pore sizes.

In eq. \ref{eq_Theory.4.3} we have derived that for lipid membranes in transitions the compressibility changes are proportional to the excess heat capacity. Therefore, we finally obtain for the permeability $P$:
\begin{equation}
\label{eq_Theory.5.6}
P=P_0+\alpha \Delta c_P
\end{equation}
where $\alpha$ is a constant. To summarize, one expects that the mean pore size and the pore number is larger when the compressibility is higher. This leads to a higher permeability. For the above calculation, it does not make a difference if a pore has twice the area or if one simultaneously finds two pores. Obviously, in the above derivation the free energy of the interface between pore wall in water has been neglected. This can only be justified by experiment. The data in Fig. \ref{Figure1b} (right) which yield a satisfactory proportionality between heat capacity and permeability suggest that in fact the free energy of a pore is dominated by the area work as assumed in eqs.\ref{eq_Theory.5.5} and \ref{eq_Theory.5.6}. This neglects the edge term of the pore which was contained in eq. \ref{eq_Theory.1}. The effect of voltage, that was heuristically included in eq. \ref{eq_Theory.1} is also not considered here because a satisfying description of the effect of voltage on the melting transition is yet missing, even though experimental evidence suggests that this effect must be significant (see also section \ref{Variables.4}). 

\section{Influence of changes of variables on the lipid channels}\label{Variables}
From the above derivations it is clear the changes in all intensive variables should have a predictable influence on the permeability when their effect on the melting transition is known. According to the list given in eq. \ref{eq_Theory.2.8} this includes temperature, hydrostatic pressure, lateral pressure, electrostatic potential (voltage) and the chemical potentials (e.g., protein, proton or calcium concentrations). In this paragraph it will be shown that all of these dependencies exist. This is especially so if the lipid membrane is close to a melting transition (or any other kind of transition). It was shown by \cite{Heimburg2005c} (findings are also reviewed in \cite{Heimburg2007a}) that biomembranes are in fact often found close to a melting transition in their membranes under physiological conditions (typically about 10-15$^{\circ}$ above the heat capacity maximum of a relatively broad transition). Therefore, any variable that potentially changes the transition will display an influence on the occurrence of lipid ion channel events. That this is in fact the case is shown below for a number of thermodynamic variables. 

\subsection{Temperature}\label{Variables.1}
The previous section justified why the formation of lipid channels is most likely in the melting transition of lipid membranes. This implies that temperature must be an important variable. Let us consider the following very simple mass action scheme for a transition at constant pressure
\begin{equation}
\label{eq_Variables1}
K=\frac{[F]}{[G]}=\exp\left(-\frac{\Delta G}{RT}\right)=\exp\left(-\frac{\Delta H-T\Delta S}{RT}\right)
\end{equation}
where $[F]$ is the concentration (or fraction) of fluid lipids and $[G]$ is the concentration of gel lipids, respectively. 
In the following we will assume that the state of the membrane and its heat capacity depends on $\Delta G$. For $\Delta G>0$ we find a larger fraction of gel phase, while for $\Delta G<0$ the membrane is mostly in its fluid phase. 
The melting temperature, $T_m$, shall be defined as the temperature where $[G]=[F]$, or $K=1$ and
\begin{eqnarray}
\label{eq_Variables1.1}
\Delta H-T_m\Delta S = 0 \quad &\rightarrow &\quad \Delta S=\frac{\Delta H}{T_m} \nonumber\\
& \rightarrow & \Delta G=\Delta H\left(\frac{T_m-T}{T_m}\right)
\end{eqnarray}
Thus, obviously a change in temperature will lead to a change of the free energy difference between gel and fluid phase. Since at the melting point, T$_m$, $\Delta G$ is zero, the fluctuations are maximum because the free energy necessary to move the system from one state into the other one is zero, and thermal noise is sufficient to do so. Therefore, the likelihood to form pores and the permeability must be strongly influences by temperature. This was shown for the macroscopic permeability experiments in section \ref{MacroscopicExp} \cite{Papahadjopoulos1973, Jansen1995, Sabra1996, Blicher2009} and for the lipid channel formation in section \ref{ChannelsExp.2} \cite{Antonov2005, Blicher2009}. \\
The effect of temperature could be expressed as being a 'thermo-sensitivity' of channel formation in pure lipid membranes.\\

\begin{figure}[htb!]
    \begin{center}
	\includegraphics[width=8cm]{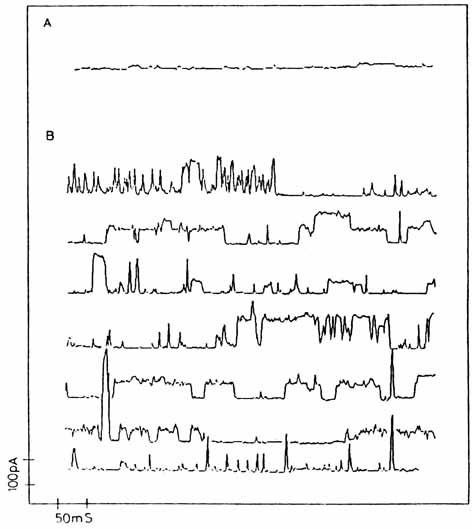}
	\parbox[c]{8cm}
{ \caption{\textit{Channel events in a membrane made of soybean phosphatidylcholine in 1M unbuffered NaCl, glass pipette tip diameter ca. 5$\mu$m, at 500 mV. The top trace (A) recorded before application of suction does not show any current fluctuations. B. The bottom traces were recorded with suction applied. The change in membrane area was monitored by recording the membrane capacitance that increased from 260 to 510 pF. From \cite{Kaufmann1989c}. }
	\label{Figure6d}}}
    \end{center}
\end{figure}
\subsection{Pressure and lateral tension}\label{Variables.2}
In the following, we will express the state of the system by the Gibbs free energy difference between fluid and gel state. We will show that it depends not only on temperature but also on all the other intensive thermodynamic variables.\\

The enthalpy is a function of pressure: $H=E+pV$. If $\Delta H_0$ is the enthalpy difference between fluid and gel phase at a pressure of 1 bar,  $\Delta V$ the respective volume difference, and  $\Delta p$ the pressure difference with respect to 1 bar, we obtain
\begin{eqnarray}
\label{eq_Variables2.1}
T_m(\Delta p) & = &\frac{\Delta H}{\Delta S}=\frac{\Delta H_0+\Delta p\Delta V}{\Delta S}\nonumber\\
&=&\frac{\Delta H_0 (1+\gamma_V \Delta p)}{\Delta S}\\
& = &T_m(1+\gamma_V\Delta p)\nonumber
\end{eqnarray}
or 
\begin{equation}
\label{eq_Variables2.2}
\Delta T_{m}(\Delta p) =\gamma_V T_m\Delta p 
\end{equation}
where $\gamma_V=7.8\cdot 10^{-19}$m$^3$/J is the parameter defined in eq. \ref{eq_Theory.4.1}. Similarly, the shift of the melting transition by lateral pressure (or tension) is
\begin{equation}
\label{eq_Variables2.3}
\Delta T_{m,} (\Delta \Pi)=\gamma_A T_m\Delta \Pi 
\end{equation}
with $\gamma_A=0.89$ m$^2$/J (see eq. \ref{eq_Theory.4.2}).
It has been shown by Kaufmann et al. \cite{Kaufmann1989c} that the application of suction on a patch pipette can generate channel events. Suction changes the lateral tension/pressure of the lipid membrane and thus has an influence on the state of the lipid membrane. Fig. \ref{Figure6d} shows suction induced channel formation in soybean phosphatidylcholine in 1M NaCl at 500mV. \\
The influence of pressure on channel appearance could be called  a 'mechano-sensitivity' of channel formation in pure lipid membranes.
 
\subsection{Anesthetics, neurotransmitters and other drugs}\label{Variables.3}
\begin{figure*}[htb!]
    \begin{center}
	\includegraphics[width=14cm]{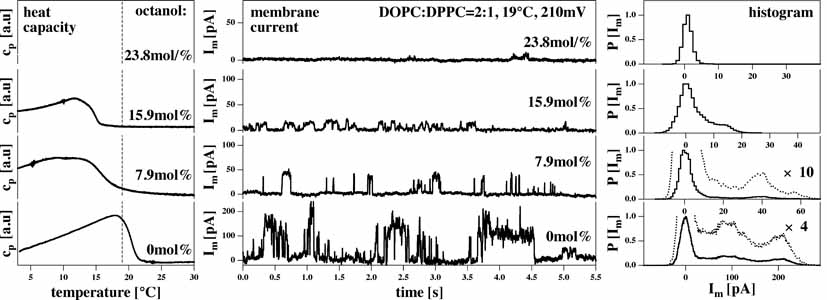}
	\parbox[c]{16cm}
{ \caption{\textit{Effect of the anesthetic octanol on the melting transition of a DOPC-DPPC=2:1 mixture in 150mM KCl at 19$^{\circ}$C. Left: Heat capacity profiles in the presence of increasing quantities of octanol demonstrate the shift of the transition caused by the anesthetic drug. The vertical dashed line indicates the experimental temperature. Center: Membrane current recorded at 210 mV demonstrates that octanol both decreases the frequency of current events and the amplitude until currents are completely 'blocked'. Right: Current histograms of the center traces. }
	\label{Figure6d2}}}
    \end{center}
\end{figure*}
Anesthetics can lower melting temperature by $\Delta T_{m,x_A}$ due to a very simple effect called melting point depression \cite{Heimburg2007c}. It is described by 
\begin{equation}
\label{eq_Variables3.1}
\Delta T_{m,x_A} =\frac{R\,T_m^2}{\Delta H}x_A
\end{equation}
where $x_A$ is the molar fraction of anesthetics dissolved in the membrane, $\Delta H$ is the melting enthalpy of the membrane, and $T_m$ is the melting temperature. To arrive at this law one has to assume that the anesthetics molecules dissolve ideally in the fluid lipid membrane, while they are insoluble in the gel membrane. The success of eq. \ref{eq_Variables3.1} in describing the effect of anesthetics on melting transition \cite{Heimburg2007c} does not only justifies this assumption but also explains the famous Meyer-Overton rule that states that the effectiveness of an anesthetic is proportional to its solubility in oil (the membrane core), independent of its chemical composition \cite{Meyer1899,Overton1901,Overton1991}. \\
Fig. \ref{Figure6d2} shows the effect of the anesthetic octanol on membranes made of a DOPC-DPPC mixture. In the absence of octanol, the experiment is performed at the heat capacity maximum where many channel events can be found. Increasing amounts of octanol shift the transition maximum towards lower temperatures and thereby lower the heat capacity at the experimental temperature. As a consequence, the frequency and amplitude of the current events decreases until channel events completely disappear. This effect is also discussed in detail in \cite{Blicher2009, Wodzinska2009}. If the membrane is recorded at a temperature below that transition, anesthetics can drastically increase the membrane permeability because they move it into the transition regime \cite{Blicher2009}.\\
Sabra et al. \cite{Sabra1996} showed that the macroscopic permeability of DMPC membranes for Co$^{2+}$ was altered by the insecticide lindane in agreement with the influence of the drug on the heat capacity profile. Other drugs as neurotransmitters also have the potential to shift melting transitions and to alter the fluctuations in membranes. This was for example shown for serotonine \cite{Seeger2007} that has a quite significant influence on melting profiles. Thus, neurotransmitters also should display the potential to influence lipid channel events considerably. More so, every drug that alters the melting behavior of a membrane can change the appearance of lipid channels. This includes proteins and peptides that can alter lipid channel behavior without being channels themselves, as discussed in section \ref{Interfaces2}. \\
The possibility to "block" or induce lipid channel events in protein-free membranes by drugs could be considered as an agonist or antagonists effect, or "channel-gating". It does not require specific binding to any macromolecule but rather represents an unspecific physicochemical process.

\subsection{Voltage, pH and calcium}\label{Variables.4}
About 10\% of the lipids in most biomembranes carry negative charges in their head groups. Charged membranes display an electrostatic potential that depends on the charge density. At typical conditions of biological membranes, i.e., low charge density (10\% of the lipids) and high ionic strength ($\approx$150mM) the electrostatic surface potential of a membrane  $\Psi_0$ can be approximated by the low potential approximation that is given by 
\begin{equation}
\label{eq_Variables4.1}
\Psi_0=\frac{1}{\epsilon\;\epsilon_0 \kappa}\sigma\propto \frac{\sigma}{\sqrt{c}}
\end{equation}
where $\epsilon_0$ is the vacuum permittivity, $\epsilon$ is the dielectric constant of water ($\approx$ 80) and $\sigma=q/A$ is the charge density. $\kappa$ is the Debye constant that is proportional to $\sqrt{c}$ with c being the ionic strength. Consequently, the larger the ionic strength and the smaller the surface charge density, the smaller the surface potential. From this the electrostatic free energy can be calculated:
\begin{equation}
\label{eq_Variables4.2}
F_{el}=\int_{\sigma'=0}^{\sigma}\Psi_0 dq \propto \frac{q^2}{A\sqrt{c}}
\end{equation}
\begin{figure}[htb!]
    \begin{center}
	\includegraphics[width=8cm]{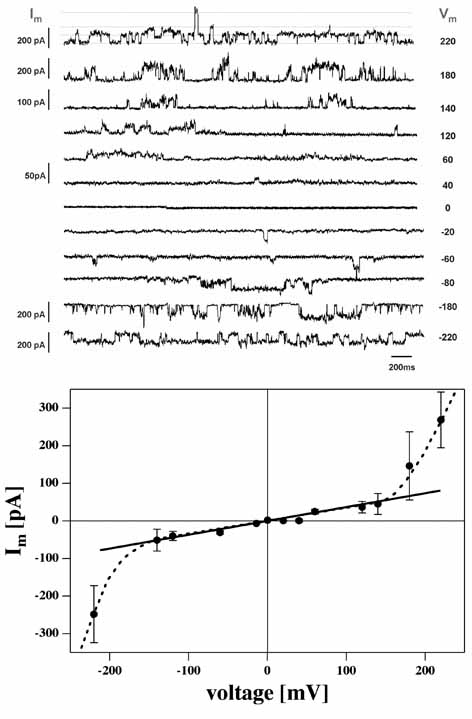}
	\parbox[c]{8cm}
{ \caption{\textit{Currents through a membrane made of DOPC:DPPC=2:1 mixture at 19$^{\circ}$ and 150mM KCl recorded at different voltages. One finds a linear current-voltage relationship for $|V_m|<$150 mV that becomes non-linear at higher voltages indicating that the voltage alters the state of the membrane. From \cite{Wodzinska2009}.}
	\label{Figure6f}}}
    \end{center}
\end{figure}
i.e., the potential is quadratic in the total surface charge and inversely proportional to the membrane area. Since biological membranes are asymmetrically charged (typically with most negative lipids being on the cytosolic side of the cell membrane \cite{Rothman1977, Rothman1977b}) one expects a transmembrane voltage of order 50mV \cite{Heimburg2007b}. 
The molecular area of the gel state of a lipid membrane is about 25\% smaller than that of a fluid lipid \cite{Heimburg1998}. This implies that the electrostatic free energy of a gel membrane is larger than that of a fluid lipid membrane because the charge density is larger in the gel state. The decrease of the electrostatic free energy contribution upon membrane melting, $\Delta F_{el, L}$, leads to a lowering of the phase transition temperature by
\begin{equation}
\label{eq_Variables4.3}
\Delta T_{m, el} =\frac{\Delta F_{el,L}}{\Delta S}
\end{equation}
where $\Delta F_{el,L}<0$ is the electrostatic free energy difference between fluid and gel lipids. The mathematical expression for $\Delta F_{el, L}$ (not given here) has been derived in \cite{Trauble1976} and been reviewed in \cite{Heimburg2007a}.  The above equation implies that the change in lipid state from gel to fluid will lower the free energy of the membrane because the membrane area increases and the potential decreases. Further, by reversing the argument it is obvious that application of voltage across a membrane must change the lipid state. In particular, charged and uncharged membranes should respond differently to voltage. Since negative lipid head groups can be protonated upon lowering the pH, the membrane state must be pH dependent.\\
\begin{figure}[hb!]
    \begin{center}
	\includegraphics[width=8cm]{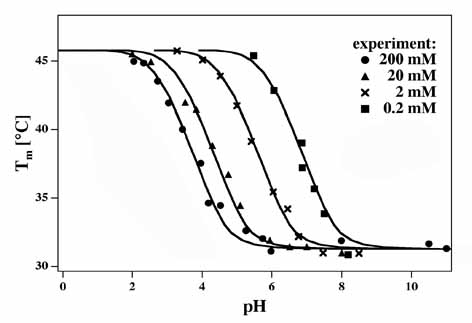}
	\parbox[c]{8cm}
{ \caption{\textit{Effect of charge on the transition temperature of dimyristoyl methylphosphatidic acid membranes as a function of pH for four different ionic strengths (adapted from \cite{Trauble1976}). The transition midpoint changes by about 14 K towards higher temperature upon protonation of the membrane. The pK$_A$ of the membrane is at higher pH values when the salt concentration is lower.}
	\label{Figure6e}}}
    \end{center}
\end{figure}
\textbf{Voltage: }The above argument showed that the lipid state should depend on transmembrane voltage. An indication for this is shown in Fig.\ref{Figure6f} that displays the transmembrane currents as a function of voltage, and the corresponding current-voltage relationship of the channels. It can be seen that the I-V curve is linear in a regime from about -150 to +150 mV (i.e., conductance is constant), while the I-V relation becomes non-linear at $|V_m| >$ 150 mV. That voltage can reversibly induce channel events was reported by \cite{Kaufmann1989c}. Antonov et al. \cite{Antonov1990} showed that voltage increases the melting transition in charged lipid membranes without significantly altering the transition temperature of zwitterionic lipids. However, in our daily lab routing we reproducibly find that zwitterionic membranes not displaying channels at low voltage show such events upon increasing the voltage (in agreement with eq. \ref{eq_Theory.1}). Therefore, we assume that voltage also has an effect on zwitterionic membranes. We further suspect that lipid ion channel formation is closely linked to the phenomenon of electroporation (cf. \cite{Neumann1999}). It should be added zwitterionic membranes are piezoelectric due to the electrical dipoles of their head groups \cite{Jakli2008}. This was also shown in monolayer experiments demonstrating different dipole potentials in the gel and the fluid phases of the membranes and an effect of voltage on the lipid state \cite{Lee1995}. \\
The influence of voltage could be called 'voltage-sensitivity' or 'voltage-gating'.\\

\begin{figure}[htb!]
    \begin{center}
	\includegraphics[width=8cm]{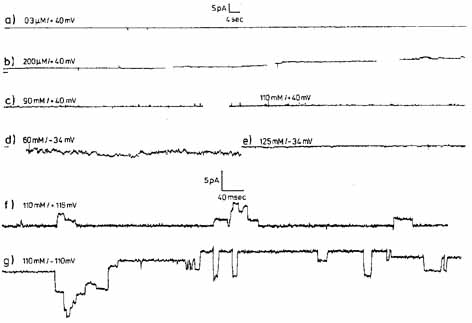}
	\parbox[c]{8cm}
{ \caption{\textit{Transmembrane currents through soybean phosphatidylcholine membranes in an unbuffered medium with 1M KCl. Traces are shown with different proton concentrations (pH range from 6.5 (trace a) to 1 (traces f and g)) at different positive and negative voltages. Transmembrane currents are induced at low pH close to the pK$_A$ of the membrane. From \cite{Kaufmann1983a}.}
	\label{Figure6g}}}
    \end{center}
\end{figure}
\textbf{Protons and pH:} Let us consider a membrane that is negatively charged at neutral pH. Upon lowering the pH the phosphate groups will be protonated. This case has been treated in much detail by Tr\"auble and collaborators \cite{Trauble1976}. They provided the electrostatic theory briefly outlined above and showed how the melting temperature depends on pH and ionic strength. Upon complete protonation the transition midpoints increases by 10-20 degrees. This is shown in Fig. \ref{Figure6e} for dimyristoyl methylphosphatidic acid (DMMPA) membranes \cite{Trauble1976}. This is shown in Fig. \ref{Figure6e}. Similarly, zwitterionic lipids my become become positively charged at low pH values around1-2. Their transition midpoint will decrease due to the electrostatic contribution of the lipid charges.
From this it must be followed that the occurrence of lipid ion channels should be pH-sensitive. This has in fact been shown by Kaufmann and Silman \cite{Kaufmann1983a}, see Fig. \ref{Figure6g}. They found that channels can be induced by protons in the pH regime around 1 that is close to the pK$_A$ of the zwitterionic soybean phosphatidylcholine membrane.\\

\begin{figure}[htb!]
    \begin{center}
	\includegraphics[width=8cm]{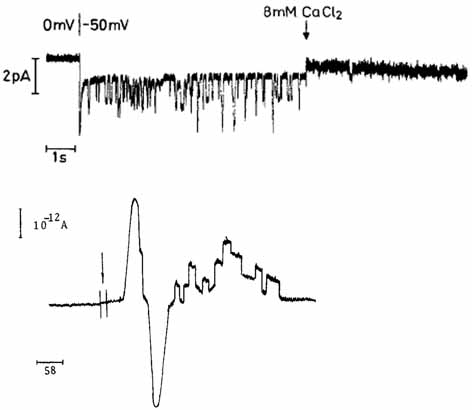}
	\parbox[c]{8cm}
{ \caption{\textit{Influence of calcium on the occurrence of lipid channels. Top: Currents through bovine brain phosphatidylserine : cholesterol=2:1 (w/w) membranes in 200mM cyclamate and pH 7.4. Addition of 8mM CaCl$_2$ inhibited the currents. From \cite{Gogelein1984}. Bottom: Generation of quantized currents in DPPC membranes by calcium at T=64$^{\circ}$C (the melting temperature of DPPC is 41$^{\circ}$C). From \cite{Antonov1985}.}
	\label{Figure6h}}}
    \end{center}
\end{figure}

\textbf{Calcium: } Due to its two positive charges, calcium also displays a pronounced effect on lipid membranes. It binds to both, negatively charged and somewhat weaker to zwitterionic membranes. This is likely to be associated with the electrostatics of the membranes. The two calcium charges probably cross-link the phosphate groups of two adjacent lipids. Typically, calcium increases melting temperatures. Depending on state of the membrane (e.g., whether found below or above a transition) calcium may therefore induce or inhibit lipid ion channels. This is shown in Fig. \ref{Figure6h}. G\"ogelein \& Koepsell \cite{Gogelein1984} showed that channel events in brain phosphatidylserine can be "blocked" or inhibited by calcium (Fig. \ref{Figure6h}, top, while Antonov et al. \cite{Antonov1985} showed the reverse effect for DPPC membranes above the transition (Fig. \ref{Figure6h}, bottom).
The effect of calcium corresponds to a 'calcium-control' or 'calcium-gating' of the lipid channels. As for the case of drugs, no specific binding site at a particular macromolecule is involved. One should bear in mind that both pH and calcium effects are related to the surface potential and thus to the voltage effects. This includes the adaptation of lipid composition reported for many bacteria under various temperature, pressure or solvent conditions (see below).

\subsection{Other variables}\label{Variables.5}
Everything that potentially changes the state of the lipid membrane will display the potential to generate or inhibit lipid ion channels. We just list a few of these:
\begin{itemize}
  \item membrane curvature, because one of the work terms in the internal energy differential is related to curvature, and curvature has an influence on the membrane phase behavior \cite{Baumgart2003}.
  \item dipolar interactions, because the work to orient a dipole (e.g., the lipid head group) in a field is part of the differential of the internal energy. It has been shown that electrostatic fields can alter the phase behavior of zwitterionic monolayers \cite{Lee1995}.
  \item hydrolysis of membranes by enzymes, because hydrolysis products change the phase behavior of the membranes \cite{Burack1993}. Any change in membrane composition may have a similar effect.
  \item membrane proteins. This case is discussed in section \ref{Interfaces2}
  \item ...
\end{itemize}

\subsection{Free energy of channel formation}\label{Variables.6}
Let us consider a membrane in a particular state that is not within the melting transition. This membrane displays a very small likelihood to display lipid ion channels. In order to increase the channel probability, the membrane has to be brought into the chain melting regime, which can be done be altering the various parameters described above. Heimburg \& Jackson \cite{Heimburg2007c} showed that the free energy that is necessary to bring the membrane into a regime where channels occur (i.e., the free energy difference between the fluid and the gel phase) is given by:
\begin{eqnarray}
\label{eq_Variables6.1}
&&\Delta G(T, x_A, \Delta p,...)=\nonumber\\
&&\Delta H\left(\underbrace{\frac{T_m-T}{T_m}}_{\mbox{\tiny{temperature}}}-\underbrace{\frac{RT}{\Delta H}x_A}_{\mbox{\tiny{anesthetics}}}+\underbrace{\gamma_V\Delta p\frac{T}{T_m}}_{\mbox{\tiny{pressure}}}+...\right)
\end{eqnarray}
For each of the thermodynamic variables, there is one term influencing the free energy difference. One could add terms for the electrostatic potential, for pH, calcium concentration, curvature, etc., which has not been done here. The content of this equation is: The further away a membrane is from its melting transition the more free energy has to be provided to generate channels. Further, for a given value of the free energy difference the likelihood of channel formation is constant, but the values of the individual variables can change. The effect of anesthetics can be compensated by hydrostatic pressure, pH, or temperature change. These effects have been documented for anesthesia \cite{Johnson1950, Punnia-Moorthy1987, Heimburg2007c}  yielding the well-known effects of pressure-reversal of anesthesia and the impossibility to anesthetize inflamed tissue with normal anesthetic doses.

\section{Channel lifetimes}\label{Life times}
It has been found in many experiments that relaxation time scales are closely coupled to the melting process in membrane \cite{Tsong1974, Gruenewald1980, Elamrani1983, Blume1986, vanOsdol1989, vanOsdol1991a, Grabitz2002, Seeger2007}.
The fluctuation-dissi{\-}pation theorem reveals a fundamental coupling between the magnitude of fluctuations (expressed in heat capacity, compressibilities, capacitance, ...) and the the relaxation time scales (indicating how fast a system relaxes to equilibrium after a perturbation), which are identical to the fluctuation time scales. Since we consider here  the lipid channel formation process basically as the consequence of a local fluctuation in state, it is obvious that open and closed life times of a channel must be intimately related to the fluctuation time scale. Basically, the fluctuation-dissipation theorem states that at regimes of high compressibility and high heat capacity, fluctuations are slow. Thus, channel life times are expected to be long in the phase transition regime when their likelihood of formation is large.

\subsection{Theoretical considerations}
We now return to the theoretical treatment of fluctuations given in sections \ref{Theory.2} and \ref{Theory.3}. On the basis of this description the fluctuation life times of membranes can be obtained \cite{Grabitz2002, Seeger2007}. Lars Onsager \cite{Onsager1931a} treated time dependent phenomena such that he assumed a  proportional relation between fluxes (rates of change of a variable) with the thermodynamic forces, an assumption corresponding to the equation of motion in viscous fluids where the influence of inertia is small. Following this approach, the flux of heat (enthalpy) of the membrane after a perturbation back to equilibrium (which could consist of the thermal collisions related to Brownian motion) is given by
\begin{equation}
\label{eq_LifeTime.1.1}
\frac{d\xi_H}{dt}=L\cdot X_H
\end{equation}
where $X_H$ is the thermodynamic force related to enthalpy differences (cf. section \ref{Theory.2}), given by $\partial S/\partial H$ with the thermodynamic variable $\xi_H=H-\left<H\right>$. $L$ is a material constant that we assume to be temperature independent. This equation is a simple version of Onsager's phenomenological equations if only one independent thermodynamic variable is present.
\begin{equation}
\label{eq_LifeTime.1.2}
X_H=\frac{\partial S}{\partial H}=-g_{HH}\cdot \left(H-\left<H\right>\right)
\end{equation}
According to eqs. \ref{eq_Theory.2.9}, \ref{eq_Theory.3.2} and \ref{eq_Theory.3.3},
\begin{equation}
\label{eq_LifeTime.1.3}
c_p=\frac{\left<H^2\right>-\left<H\right>^2}{kT^2}=\frac{1}{T^2 g_{HH}}\quad\mbox{or}\quad g_{HH}=\frac{1}{T^2 c_p}
\end{equation}
This leads to
\begin{equation}
\label{eq_LifeTime.1.4}
\frac{d(H-\left<H\right>)}{dt}=-\frac{L}{T^2 c_p}\left(H-\left<H\right>\right)
\end{equation}
with the solution
\begin{eqnarray}
(H-\left<H\right>)(t) & = & (H-\left<H\right>)_0\cdot\exp\left(-\frac{L}{T^2 c_p}\;t\right) \nonumber\\
 & \equiv &  (H-\left<H\right>)_0\cdot\exp\left(-\frac{t}{\tau}\right)
\end{eqnarray}
introducing the relaxation time
\begin{equation}
\label{eq_LifeTime.1.6}
\tau=\frac{T^2 c_p}{L}
\end{equation} 
This equation implies that the relaxation (or fluctuation) time scale is large when the heat capacity is high. If we consider the channel formation as a fluctuation, we have to conclude that the typical time scale of channel opening and closing is proportional to the heat capacity. In particular, in the transition where the likelihood of lipid ion channel formation is high, they should also display long open times.

It has been found that for pure lipid membranes the phenomenological coefficient $L$ is of order 6-20$\cdot$10$^8$ J K/mol sec. For a DMPC multilayer one finds a relaxation time of about 30 seconds at the maximum. Transitions in biological membranes are much broader and the heat capacity maximum displays much smaller values. Assuming the same phenomenological coefficient as for DMPC membranes, one estimates about 100 milliseconds at the $c_p$ maximum of bovine lung surfactant or \textit{E.coli} membranes. Since biological membranes display a $c_p$ maximum some degrees below body temperature, the expected open times at physiological temperature is expected to be of the order of a few milliseconds. Interestingly, this time scale is just of the same order as the dwell times of  quantized current events in section \ref{ChannelsExp}.\\

Summarizing: While the channel open times in artificial membranes at the $c_p$ maximum may be very long (of order seconds), they are expected to be of millisecond order for biomembranes at physiological temperature. This time scale, expected for lipid channels, is interestingly also exactly that found for protein channels. This is unlikely to be accidental.

\begin{figure}[tb!]
    \begin{center}
	\includegraphics[width=7cm]{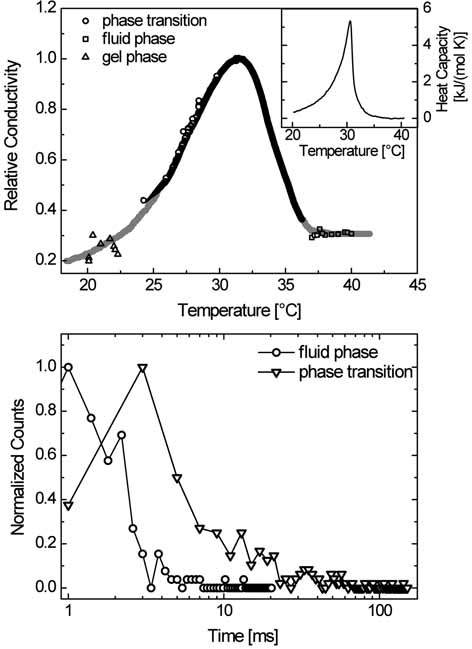}
	\parbox[c]{8cm}
{ \caption{\textit{Top: Relative conductance of a D15PC-DOPC (95/5) mixture as a function of temperature at V$_m$ = 90mV. The insert shows the melting profile of the lipid mixture. Adapted from \cite{Wunderlich2009}. Bottom: Logarithmic plot of the timescales of the current fluctuation in the fluid phase (circles) and the phase transition regime (triangles).  The average lifetimes center around 3ms in the fluid and 20ms in the phase transition, which is in good agreement with our theoretical prediction. From \cite{Wunderlich2009}.}
	\label{Figure7a}}}
    \end{center}
\end{figure}
\subsection{Experiments}
It has been shown by \cite{Wunderlich2009} that channel life times are in fact much longer in the transition of a lipid mixture that displays a transition half width comparable to that of biomembranes. This is shown  in Fig.\ref{Figure7a}. Both heat capacity profile and mean conductance of a lipid mixture of DC$_{15}$PC and DOPC (95:5) display maxima at about 30$^{\circ}$C (Fig.\ref{Figure7a}, top). The typical open time of the lipid channels is of order 20 milliseconds, while it is rather 3ms above the transition.  If one compares the time scales of the current fluctuations in the phase transition regime and in the fluid phase one finds that they are about one order of magnitude different (Fig.\ref{Figure7a}, bottom), in approximate agreement with the above calculation (eq. \ref{eq_LifeTime.1.6}). The timescale in the fluid phase follows a single-exponential decay while in the transition it can only be approximated by a double-exponential function. Gallaher et al. \cite{Gallaher2009} provided an analysis showing that the current fluctuation time scales in the transition regime obey a power law rather than single exponential behavior as suggested by the derivation above. The authors rationalized this by assuming that the conduction pores may exist in different states with different gel-fluid environments. Figs. \ref{Figure5a} and \ref{Figure5b} also suggest that there might be more than one possible pore geometry with similar pore dimension. \\
Summarizing, it seems as if one can understand the time scale of lipid ion channel opening and closing with simple non-equilibrium thermodynamics (fluctuation-dissipation). This means that heat capacity, phase behavior and time scales are different aspects of the same physics. Whenever heat capacities and permeabilities are high, time scales are larger. However, while the overall change in time scales is consistent with the above derivation, multi-exponential or power law behavior is not contained and requires more sophisticated analysis. \\
Heat capacity profiles can be influenced by anesthetics, neurotransmitters and by membrane peptides. Interestingly, it has been shown by \cite{Seeger2007} that those $c_p$ changes are accompanied by equivalent changes in the relaxation time scales, such that those molecules cause a predictable change of the relaxation time scales. The role of proteins is outlined in more detail in the following section.


\section{Interfaces: Domains and proteins}\label{Interfaces}
Fluctuations are large in regions where the free energy of gel and fluid lipids are similar. Therefore, they are maximum at the melting transition. As shown above, many variables can be adjusted (pressure, temperature, ...) such that the membrane system is situated in this regime. However - such fluctuations must not be the same everywhere in the membrane if the membrane is heterogenous. Biological membranes are complex mixtures of lipids with different melting points and different hydrophobic lengths, and many other molecules including cholesterol and proteins. These membranes display phase separation and domain formation phenomena. Therefore, one must expect that also the fluctuations display different magnitude in different membrane regions. It is especially interesting to consider interfaces between fluid and gel domains. \\
The magnitude of fluctuations depends on the length scale over which they are monitored. One can consider fluctuations of the whole membrane, which enter into the heat capacity or the macroscopic compressibility. If one considers the fluctuations in windows of smaller size, e.g. on the length scale of individual domains or on molecular scale, one obtains local heat capacities and compressibilities. There are different ways of probing the local fluctuations, e.g., 
\begin{itemize}
  \item experimentally:  One can investigate the softness of a membrane by AFM. One obtains the compressibility on the length scale of an AFM-tip (around 10nm). This has been done for peptide containing membranes in \cite{Oliynyk2007}. 
  \item theoretically by simulations: In Monte Carlo simulation, one can directly obtain the fluctuation on arbitrary scales. Such simulations typically use experimental observables such as melting enthalpy and entropy, and the shape of the melting profile in order to determine the simulation parameters (e.g., \cite{Mouritsen1990, Mouritsen1995a, Mouritsen1995b, Heimburg1996a, Ivanova2003}). In principle, such information is also contained in molecular dynamics (MD) simulations. However, to our knowledge MD has not been used yet to obtain such information.
\end{itemize}

\begin{figure}[htb!]
    \begin{center}
	\includegraphics[width=8cm]{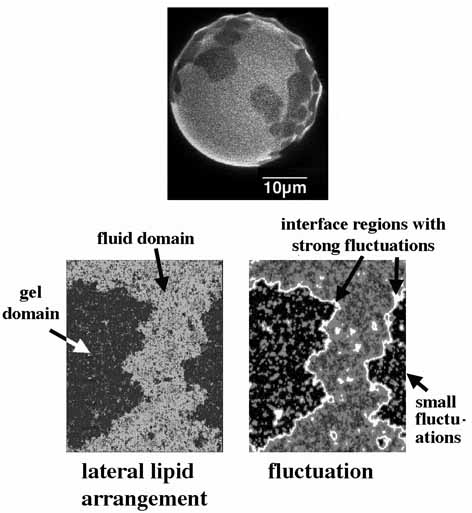}
	\parbox[c]{8cm}
{ \caption{\textit{Top: Confocal microscopy image of a DLPC-DPPC (1:2) mixture in its melting regime at 27$^{\circ}$C. Dark regions represent gel domains, bright regions are fluid \cite{Fidorra2004}. Bottom, left: Snapshot of a Monte Carlo simulation of a DMPC-DSPC (1:1) mixture in the melting regime (36.9$^{\circ}$C). Dark shades correspond to gel domains while bright shades represent fluid domains. Bottom, right: Local fluctuations of the same matrix. Bright shades correspond to large fluctuations while dark shades correspond to small fluctuations. The magnitude of the fluctuations is related to the membrane permeability and the likelihood of lipid channel formation. For the example shown, fluctuations are larger in the fluid regions than in the gel regions, but are maximum at the interface between the domains. These regions are those of maximum permeability. From \cite{Seeger2005}.}
	\label{Figure8d2}}}
    \end{center}
\end{figure}
\subsection{Domain interfaces}\label{Interfaces1}
In Fig. \ref{Figure8d2} domain formation of lipid mixtures is shown. Confocal microscopy images of large unilamellar vesicles display domains of gel and fluid state (Fig. \ref{Figure8d2}, top). Such domain formation can be described by lattice Monte Carlo simulations (Fig. \ref{Figure8d2}, bottom left). Details are, e.g., given in \cite{Seeger2005}. For the purpose of this review only the results are important. The fluctuations of single sites in this matrix (local fluctuations) are shown in Fig. \ref{Figure8d2} (bottom right). It is found that under the conditions indicated in the figure legend the gel domains display much smaller local fluctuations than the fluid domains. The largest fluctuations, however, are seen at the domain interface. Thus, while the domain interfaces are the regions of largest permeability, also the domains themselves contribute to permeation (in Fig.\ref{Figure8d2} (bottom right) especially the fluid domains). This sheds further light on the two possible explanations for the location of permeation pores discussed in section \ref{Theory.5}.

\subsection{Proteins as membrane perturbations}\label{Interfaces2}
\begin{figure}[b!]
    \begin{center}
	\includegraphics[width=9cm]{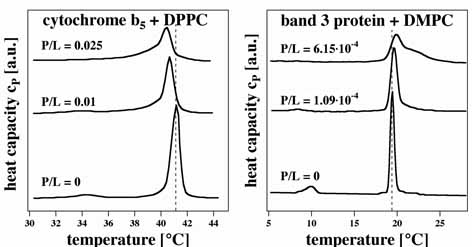}
	\parbox[c]{8cm}
{ \caption{\textit{Influence of proteins on the melting behavior of lipid membranes. Left: Cytochrome b$_5$ lowers the melting point (from \cite{Heimburg2007a} adapted from \cite{Freire1983}). Right: Band 3 protein of erythrocytes increases the melting point (from \cite{Heimburg2007a} adapted from \cite{Morrow1986}).}
	\label{Figure8c}}}
    \end{center}
\end{figure}
The gel-fluid domain interface is not the only region of large fluctuations. Macromolecules as proteins can also create interfaces with lipids that influence their phase behavior. Two examples are given in Fig. \ref{Figure8c}, showing the influence of cytochrome b$_5$ and band 3 protein of erythrocytes on the melting behavior of DPPC and DMPC membranes. While cyt. b$_5$ lowers the melting point, band 3 protein is increasing it. It has been shown in \cite{Ivanova2003} that this can be understood if one assumes that the first protein mixes well with fluid membranes but not with gel membranes, while the second protein mixes better with gel and not so well with fluid regions. Mouritsen \& Bloom \cite{Mouritsen1984} proposed that this is due to hydrophobic matching that compares the length of the hydrophobic core of the integral protein with the thickness of the membrane. Good miscibility suggests that the hydrophobic core of the protein has a similar length then the surrounding membrane. Bad miscibility indicates hydrophobic mismatch.\\

\begin{figure}[b!]
    \begin{center}
	\includegraphics[width=7cm]{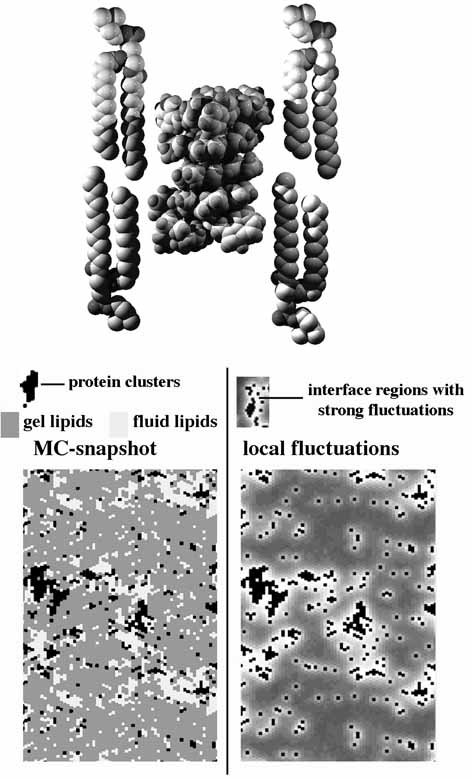}
	\parbox[c]{8cm}
{ \caption{\textit{Simulation of the antibiotic peptide gramicidin A in a DPPC membrane. Top: Gramicidin A is very short as compared to the gel lipids of DPPC and displays an unfavorable hydrophobic matching. Bottom, left: Monte Carlo snapshot of a simulation of gramicidin A in a DPPC membrane. Slightly below the melting transition one finds the membrane mostly in the gel state (gel lipids - dark grey) with some remaining fluid domains present (fluid lipids - light grey). Due to the unfavorable hydrophobic matching, the proteins aggregate in the gel phase into clusters (black dots - gramicidin molecules). Bottom, right: For the same snapshot the magnitude of the fluctuations is shown locally. Dark grey shades correspond to small fluctuations while bright shades represent large fluctuations with a high likelihood of pore formation.}
	\label{Figure8d}}}
    \end{center}
\end{figure}

Gramicidin A is a small dimeric channel peptide. As shown in Fig. \ref{Figure8d} (top) it is significantly shorter than a gel lipid of DPPC. Thus, one expects unfavorable hydrophobic matching in the gel phase of DPPC. Fig. \ref{Figure8d} (botto left) shows the distribution of gel and fluid state lipids and the proteins as obtained from a Monte Carlo simulation that was based on heat capacity analysis \cite{Ivanova2003}. The simulation was performed slightly below the melting temperature of DPPC. While most of the matrix is found in its gel state with some remaining fluid domains, the proteins aggregate into smaller clusters. Fig. \ref{Figure8d} (bottom right) shows the corresponding local fluctuations. One finds that the fluctuations are maximum at the interface of the protein clusters. This indicates that the protein interface also forms a membrane region with larger likelihood of lipid channel formation. 
\begin{figure}[htb!]
    \begin{center}
	\includegraphics[width=8cm]{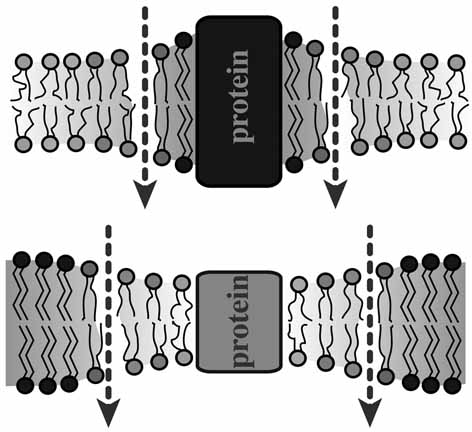}
	\parbox[c]{8cm}
{ \caption{\textit{Schematic drawing of proteins as perturbation of the lipid membrane state. Top: Protein with long hydrophobic length embedded into a fluid lipid membrane. Bottom: Protein with short hydrophobic length embedded into a gel lipid membrane. Due to wetting of the protein interface the lipids close to the protein are in a transition state with high fluctuations and an increased likelihood of lipid channel formation.}
	\label{Figure8e}}}
    \end{center}
\end{figure}
The results from the simulation are summarized in Fig. \ref{Figure8e} that is a variation of a picture from Mouritsen \& Bloom used to describe their mattress model \cite{Mouritsen1984}. If a protein with large hydrophobic length is penetrating a fluid membrane with smaller thickness, the lipids at the interface of the protein will be close to the melting regime, which is due to wetting of the hydrophobic core of the protein (Fig. \ref{Figure8e}, top). Thus, at this interface fluctuations are large and the likelihood of lipid channel formation is maximum. Similarly, if the protein has a short hydrophobic length and is surrounded by a gel membrane of larger thickness, the protein interface will constitute a location for lipid channel formation (Fig. \ref{Figure8e}, bottom). It is a general predictions from such simulations that the tendency of proteins to aggregate and their potential to trigger lipid ion channel formation are closely coupled.\\
The role of the protein in such a scenario can be seen as doping the membrane and influence the physics of its surrounding, much like in semiconductors where the doping agent gallium can influence the physics of silicon.  However, the proteins in this situation do not conduct ions themselves. Rather is the outer interface of the protein the site of permeation. All arguments arguing in favor of a selectivity\footnote{According to Einstein \cite{Einstein1902} an all-or-nothing selectivity of a membrane (or a membrane protein) for a particular substance violates the laws of thermodynamics. Therefore, we rather refer to different permeation rates.} of a protein pore (channel protein) would also be true for the outer interface. Thus, the protein in the above situation could be considered as an inverted pore (the pore interface is on the outside). The physics, however, is that of the lipid membrane.

\section{Discussion}\label{Discussion}
We have reviewed here some of the literature indicating the presence of quantized channel events in lipid membranes. It is a striking fact that one finds such events in membranes that do not contain proteins. The currents display typical amplitudes in the 10pA regime at voltages around 100mV. Typical life times are of order 10 milliseconds but can be much longer in some experiments. These currents can be inhibited by anesthetics  \cite{Blicher2009} but most likely by everything that shifts melting transitions, including peptides and neurotransmitters. They respond to pressure, tension, pH, calcium, voltage and other variables. In some sense one could consider the overall membrane as a sensitive receptor for all of these variables if one is close to a transition.

\subsection{Comparison with proteins}
When comparing the lipid ion channels with protein channels it has to be stated that 
\begin{itemize}
  \item typical conductances and lifetimes of protein and lipid channels are of similar magnitude. The two classes of events are seemingly indistinguishable.
\end{itemize}
If two events cannot be distinguished one has to reconsider what the factual evidence actually is that allows to attribute the current events to different processes. The findings on lipid channels reported here imply that\begin{enumerate}
  \item the observation of quantized currents is no proof for the action of ion channel proteins. 
  \item the action of a drug on a conduction event is no proof for specific binding of the drug to a channel protein.
\end{enumerate}
In logics, elementary rules of inference distinguish between necessary and sufficient criteria. For channel proteins it must be concluded that the finding of quantized currents alone is neither necessary nor sufficient to infer the action of a receptor. The quantized current could be though the lipid membrane, and there is no particular reason why the a quantization should indicate the action of a macromolecule. Similarly, the action of drugs is not sufficient to infer specific binding. The drug could have an effect without binding, as shown in section \ref{Variables.3} for the anesthetics. For example, the inference that the abolition of channel events in biomembranes by tetrodotoxin (TTX) proves the action of a sodium channel protein and the specific binding of TTX to this channel is not admissible without additional information, which is typically not provided in the literature.\\

The experiments reviewed in this paper do not give an indication for that protein channels do not do exactly what has been suggested in the protein ion channel literature. However, the data provide considerable evidence that one can find the same events in the absence of proteins and therefore one does not necessarily require proteins as an explanation.  While the existence of lipid channels is easy to prove because a lipid membrane can be measured in the absence of proteins, it is presently impossible to measure protein channel conductances in the absence of a membrane. Thus, it is much more difficult to actually prove that a protein ion channel is really active in an experiment. From a patch-clamp experiment it is impossible to conclude, which path a particular current event has taken. The patch area is between 1-100 $\mu$m$^2$ while the cross section of a single channel protein, e.g. the potassium channel, is rather 25 nm$^2 = 25\cdot 10^{-6} \mu$m$^2$. A typical aqueous pore has a diameter of 0.5 nm. Since no experiment can actually monitor the path of the ions and therefore demonstrate the existence of a particular pore inside a membrane protein, the argument is typically of rather indirect nature:
\begin{itemize}
  \item the protein increases the probability to find conduction events, and conductivities are small in the absence of the proteins
  \item the currents can be blocked by specific drugs as tetrodotoxin (TTX) or tetraethylamonium (TEA)
  \item site specific mutation can alter or block the currents
\end{itemize}
One must state that it has been incorrectly assumed that the lipid membrane is inert and does not display quantized conduction events. Second, proteins can act as perturbations of the membrane and thus have the potential of altering the state of the membrane. A direct proof for this effect is still missing, but it is straight-forward to conclude this from the thermodynamics described in this article. Therefore, a conductance that displays a protein concentration dependence must be taken as a proof for a correlation of the proteins with the conduction events but not as a proof for the conductance through a pore in the protein itself. Conduction events in lipid membranes depend also on the concentration of drugs like anesthetics (as shown in section \ref{Variables.3}) even though anesthetic molecules do not form channels. 
The dangers of confusing lipid with protein channels may be evident in a publication by Cannon et al. \cite{Cannon2003}. These authors found that calcium channels reconstituted in lipid mixtures are not only most active at the phase boundaries of the lipid phase diagram but also display the longest dwell times. While there is no particular reason why a protein should behave this way, this is exactly the behavior expected for lipid ion channels in lipid membranes at phase boundaries (dwell times are long in the transition, see section \ref{Life times}). The channel conductances in this paper were the same as, e.g., in the lipid mixture shown in Fig. \ref{Figure3g}. Further, MacKinnon and collaborators \cite{Schmidt2006} showed that the potassium channel conduction properties depend very sensitively on the lipid environment proving a connection between state of the lipid membrane and protein channels. The authors assumed that the voltage sensors of these proteins are lipid-controlled. The effect of anesthetics on lipid membrane pores is very similar to that reported for the acetylcholine receptor and many other channel proteins (reviewed in \cite{Wodzinska2009}.\\

\subsection{Nerve pulses}
In a series of recent publications it has been shown that close to melting transitions of lipid membranes one expects the possibility of electromechanical pulse propagation (solitons) in cylindrical membrane tubes \cite{Heimburg2005c, Lautrup2005, Heimburg2007b, Andersen2009}. These solitons in fact share many properties with the action potential in nerves that cannot be explained by the Hodgkin-Huxley model, e.g., reversible heating of the nerve, and mechanical alterations like thickness changes and length changes. Thus, it may be possible that the nerve pulse can be explained without involving ion channel proteins. However, exactly under the conditions under which such electromechanical solitons can propagate one also expects the formation of lipid ion channels. As shown in this article, everything that alters the the phase behavior will alter the lipid channel appearance and simultaneously the free energy necessary to induce a nervous pulse when following an electromechanical approach to rationalize the phenomenology of the action potential \cite{Heimburg2005c}. This is a beautiful correlation purely based on the thermodynamics of the lipid membrane that is well understood as shown here.

\subsection{Conclusion}
Taking the extreme relevance of the finding of quantized currents in artificial membrane where any effect of proteins can be excluded it is quite surprising that this phenomenon has not been studied more intensively. Not only do these currents display similar conductances as proteins channels, but obviously similar fluctuation life times. Since many biomembranes exist in a state close to melting transitions of their membranes, the striking similarity between the lipid and protein events suggests that these events are difficult or even impossible to distinguish. Should in fact some of the reported protein conductances be due to lipid pores, the theoretical description of such channels would rather lie in the thermodynamics of the membrane and its cooperative phase behavior than in the geometry of individual proteins. A discussion of the lipid channel phenomenon and its connection with protein channels is overdue.\\
Even though there have been reports on lipid ion channels for 30 years the investigation of lipid ion channels is still in an early stage. Many of the thermodynamics couplings described in this article have not been tested yet by experiment. However, already now many of the implications of applying the fluctuation-dissipation theorem to lipid channels found experimental verification, e.g., the coupling between life times and heat capacity, the effect of pH, anesthetic drugs and lateral tension. Many further experiments are needed to get better statistics of the membrane conductance and the influence of various parameters. It would in particular be nice to demonstrate the influence of proteins that cannot be channel proteins (for instance peripheral proteins) on the quantized conductance of lipid membranes. Such experiments are not difficult and will most likely be done in the near future. \\

\noindent\textbf{Acknowledgments:} 
I thank Matthias Fidorra, Katarzyna Wodzinska, Andreas Blicher, Katrine R. Laub and Klaus B. Pedersen who performed permeability measurements in my lab.  \\

\small{

}
\end{document}